\documentclass{article}

\usepackage{microtype}
\usepackage{graphicx}
\usepackage{subcaption}
\usepackage{booktabs} % for professional tables

\usepackage{hyperref}

\usepackage[accepted]{icml2026}

\usepackage{amsmath}
\usepackage{amssymb}
\usepackage{mathtools}
\usepackage{amsthm}

\usepackage[capitalize,noabbrev]{cleveref}

\theoremstyle{plain}

\theoremstyle{definition}

\theoremstyle{remark}

\usepackage[textsize=tiny]{todonotes}

\usepackage[utf8]{inputenc} % allow utf-8 input
\usepackage[T1]{fontenc}    % use 8-bit T1 fonts
\usepackage{url}            % simple URL typesetting
\usepackage{booktabs}       % professional-quality tables
\usepackage{amsfonts}       % blackboard math symbols
\usepackage{nicefrac}       % compact symbols for 1/2, etc.
\usepackage{microtype}      % microtypography
\usepackage{xcolor}         % colors
\usepackage{subcaption}    % for subfigure environment
\usepackage{graphicx}
\usepackage{appendix}

\usepackage{placeins} % for \FloatBarrier
\usepackage{comment}
\usepackage{hyperref}       % hyperlinks
\icmltitlerunning{Decoding Alignment without Encoding Alignment: A Missing Component for Interpretability}

\begin{document}

\twocolumn[
  \icmltitle{Decoding Alignment without Encoding Alignment: \\A Missing Component for Interpretability}



  \icmlsetsymbol{equal}{*}

  \begin{icmlauthorlist}
    \icmlauthor{Johannes Bertram}{tue,ellis,aic}
    \icmlauthor{Luciano Dyballa}{mad}
    \icmlauthor{T. Anderson Keller}{har}
    \icmlauthor{Savik Kinger}{yale}
    \icmlauthor{Steven W. Zucker}{yale,bio,wu}
  \end{icmlauthorlist}

  \icmlaffiliation{tue}{University of Tübingen, Germany}
  \icmlaffiliation{ellis}{ELLIS Institute Tübingen, Germany}
  \icmlaffiliation{aic}{Tübingen AI Center, Germany}
  \icmlaffiliation{tue}{University of Tübingen, Germany}
  \icmlaffiliation{mad}{School of Science \& Technology, IE University, Madrid, Spain}
  \icmlaffiliation{yale}{Department of Computer Science, Yale, New Haven CT, USA}
  \icmlaffiliation{bio}{Department of Biomedical Engineering, Yale, New Haven CT, USA}
  \icmlaffiliation{wu}{Wu Tsai Institute, Yale, New Haven CT, USA}
  \icmlaffiliation{har}{The Kempner Institute for Natural and Artificial Intelligence, Harvard University, Cambridge MA, USA}

  \icmlcorrespondingauthor{Johannes Bertram}{jb@w3a.de}

  \icmlkeywords{Machine Learning, ICML}

  \vskip 0.3in
]



\printAffiliationsAndNotice{Presented at the Mechanistic Interpretability Workshop at ICML 2026.}  % no special notice (required even if empty)

\begin{abstract}
    \looseness=-1
    RSA and CKA are standard metrics for comparing neural representations across brain regions, organisms, and deep learning models. We demonstrate a fundamental weakness: these decoding-based metrics are insensitive to \emph{encoding manifold topology} --- the internal functional organization of a neural population. In a controlled MNIST experiment, RSA, CKA, and Procrustes $R^2$ remain statistically unchanged when encoding topology is causally manipulated via an auxiliary clustering loss, while the two model populations differ significantly in attribution patterns, weight-graph assortativity, and out-of-distribution robustness. Across biological systems and machine learning models, similar decoding behavior can arise from small, non-representative subpopulations, and alignment metrics are insensitive to encoding manifold topology even when it is fundamentally altered. These findings bear directly on mechanistic interpretability: standard alignment metrics cannot distinguish whether two networks share the same computational circuits or merely produce indistinguishable aggregate outputs. We propose encoding manifolds and Gromov--Wasserstein distance as complementary diagnostics for any decoding-based similarity claim, and provide a \href{https://johannesbertram.github.io/FNN_Manifolds/index.html}{Neural Manifold Explorer} tool for further experimentation.
\end{abstract}

\section{Introduction}
\label{sec:intro}

\looseness=-1
A central goal in interpretability and neuroscience is to compare computations across neural systems --- brain regions, organisms, or deep learning models \citep{kriegeskorteRepresentationalSimilarityAnalysis2008, khaligh-razaviDeepSupervisedNot2014, schrimpfBrainScoreWhichArtificial2020, cadieuDeepNeuralNetworks2014}. The dominant approach records population responses to shared stimuli and compares them via scalar alignment scores such as RSA \citep{kriegeskorteRepresentationalSimilarityAnalysis2008} and CKA \citep{kornblithSimilarityNeuralNetwork2019}. While useful for decoding, scores over states offer only a partial view of computational organization: Two systems that are not ``digital twins'' can achieve indistinguishable alignment scores --- yet the same population state can be produced by qualitatively different internal functional organizations. The same challenge is central to mechanistic interpretability \citep{Ameisenetal2025, Lindseyetal2024}: two models trained differently, pruned, or fine-tuned may match on RSA or CKA while implementing behavior through entirely different internal circuits.

\begin{figure*}[t]
    \centering
    \includegraphics[width=\linewidth]{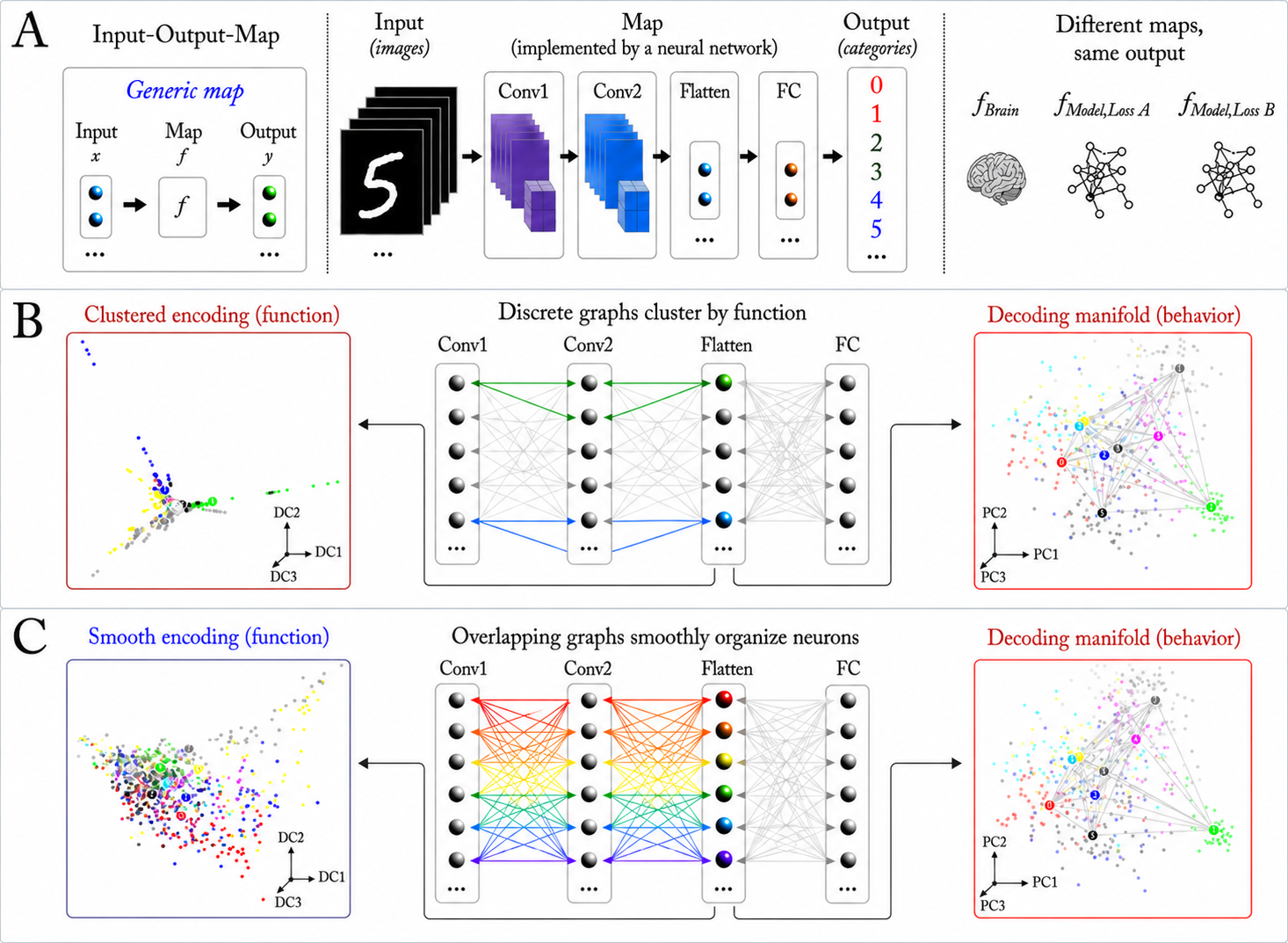}
    \caption{{\bf Encoding and decoding manifolds provide complementary views of neural systems}. (A) A neural network as an input-output map. The same behavior -- categorization -- can be achieved with different maps. (B) Stimuli can be represented in neural coordinates (decoding manifold, right, colors represent different stimuli) or neurons can be represented in stimulus coordinates (encoding manifold, left, colors represent neuron's preferred stimulus). A clustered decoding manifold implies discrimination between stimulus classes; a clustered encoding manifold (non-overlapping preferred stimulus 'arms') implies separation between neuron's response patterns. (C) A similar decoding topology (compare with B) can be achieved by a different network, with neurons that exhibit a continuous encoding topology (left), which suggests a less sparse computational graph. This artificial example is developed in Sec.~\ref{sec:mnist}.}
    \vspace{-4mm}
    \label{fig:overview}
\end{figure*}

\looseness=-1
The reason is that the map from stimuli to behavior via population states is underspecified: many different internal organizations can produce the same population-level distance geometry \citep{ganguli2012compressed, bertramHowNeuralNeural2026, ghoshWhyAllRoads2025, nellenLearningClusterNeuronal2025}. To claim functional similarity between two neural systems, one must look beyond the aggregate state and examine \emph{how} neurons (or units) are internally organized. We do so using the \emph{encoding manifold} \citep{dyballaPopulationEncodingStimulus2024, posani2025rarely}: rather than representing stimuli as points in neural-response space, the encoding manifold represents \emph{neurons} as points in stimulus-response space. The global topology of this neural point cloud --- whether it forms discrete clusters, as in retinal ganglion cell types, or a smooth continuum, as in mouse V1 --- directly reflects the functional architecture of the population, independently of any downstream readout (Figure~\ref{fig:overview}).

Decoding metrics treat the neural population as a monolithic input/output unit and return a single scalar. This paper reveals a key structural limitation: \textit{the same scalar summary of decoding behavior (RSA, CKA, or classification accuracy) can be achieved by functionally heterogeneous populations}. \\
\\
Specifically: 

\vspace{-1mm}
\begin{enumerate}
    \item A small, selective subset of neurons can replicate the full population's decoding performance and may achieve this through \textit{different} neural subpopulations.
    \vspace{-1mm}
    \item Representation alignment metrics, which operate on the same population-level distance matrices, inherit this blind spot.
    \vspace{-1mm}
    \item Encoding and decoding manifold topology can be independently manipulated by design, providing \emph{causal} rather than merely correlational evidence that decoding metrics are insensitive to internal functional organization.
\end{enumerate}
\vspace{-1mm}

\begin{figure*}[t]
    \centering
    \includegraphics[width=\linewidth]{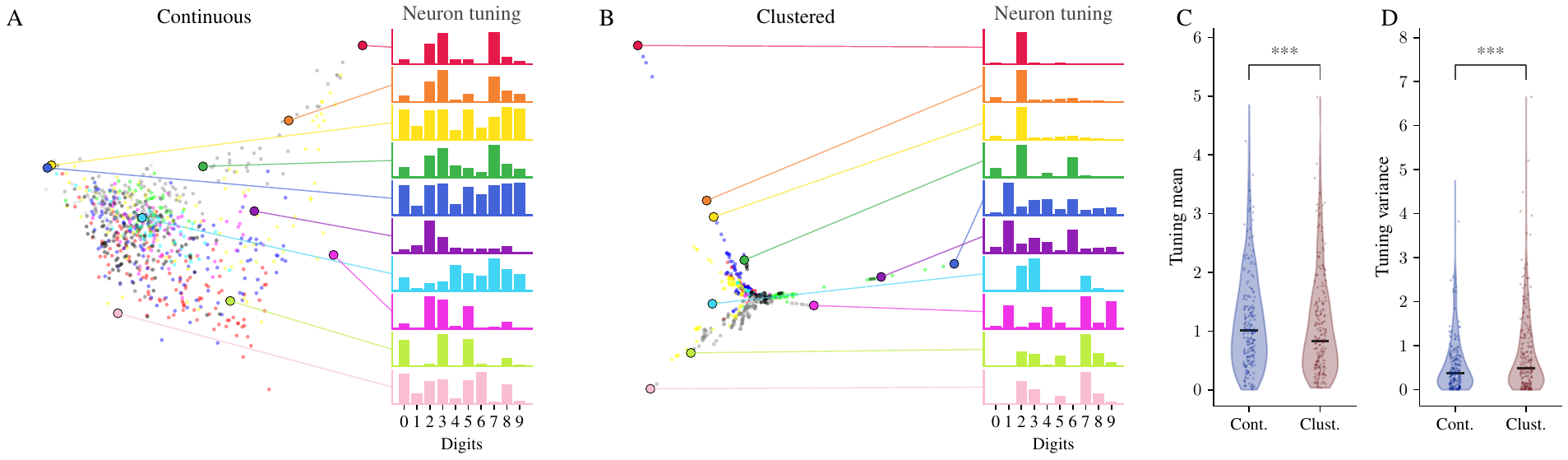}
    \caption{MNIST experiment: continuous (A) vs.\ clustered (B) encoding manifold. Tuning histograms show graded, multi-class responses in A and sharp, single-class peaks in B. Neurons in the clustered model have significantly lower tuning mean (C) and higher variance (D), indicating sharper class-selective responses. (Wilcoxon rank-sum; * $p{<}0.05$, ** $p{<}0.01$, *** $p{<}0.001$)}
    \label{fig:mnist}

\end{figure*}

\begin{table*}[h!]

\centering
\small
\caption{
\looseness=-1
MNIST results ($n{=}20$ seeds, mean\,$\pm$\,SEM).
\textbf{Top:} decoding and encoding metrics for $\lambda{=}10$ vs.\ baseline;
cross-condition metrics relative to baseline.
\textbf{Bottom:} functional organization metrics for $\lambda{=}10$ vs.\ baseline
(Wilcoxon rank-sum; * $p{<}0.05$, *** $p{<}0.001$, with Bonferroni-Holm correction.)}
\label{tab:mnist_main}
\begin{tabular}{l r r c}
\toprule
Metric & Baseline ($\lambda{=}0$) & Clustered ($\lambda{=}10$) & Sig. \\
\midrule
RSA                              & $0.958 \pm 0.001$ & $0.961 \pm 0.002$ & ns\\
CKA                              & $0.976 \pm 0.001$ & $0.976 \pm 0.001$ & ns\\
Proc.\ $R^2$                     & $0.950 \pm 0.002$ & $0.941 \pm 0.008$ & ns\\
\textbf{GW Sim (ours)}           & \textbf{0.958} $\pm 0.001$  & \textbf{0.502} $\pm 0.022$ & ***\\
\midrule
\multicolumn{3}{l}{\textit{Functional Differences}} &  \\
\midrule
Blur robustness, mean acc.\ ($\sigma_{\mathrm{blur}} \geq 4$) & $0.318 \pm 0.010$ & $0.282 \pm 0.013$ & * \\
FC Attribution Entropy           & $5.919 \pm 0.008$ & $5.756 \pm 0.007$ & *** \\
FC Attribution Overlap           & $0.463 \pm 0.002$ & $0.407 \pm 0.005$ & *** \\
FC Assortativity                 & $0.077 \pm 0.001$ & $0.086 \pm 0.002$ & * \\
Conv2 Attribution Entropy        & $3.232 \pm 0.008$ & $3.097 \pm 0.019$ & *** \\
Conv2 Attribution Overlap        & $0.902 \pm 0.003$ & $0.889 \pm 0.006$ & ns \\
Conv2 Assortativity              & $0.277 \pm 0.035$ & $0.501 \pm 0.003$ & *** \\
\bottomrule
\end{tabular}
\vspace{-2mm}
\end{table*}

\looseness=-1
We demonstrate these findings across mouse retina and V1, five Allen cortical areas, and several machine learning models (Appendix~\ref{app:data}). A model could be declared a faithful replica of cortex based on matched RSA or CKA while its internal organization is structurally distinct. We therefore propose encoding manifolds and Gromov-Wasserstein distance as complementary diagnostics \citep{dyballaPopulationEncodingStimulus2024, bertramHowNeuralNeural2026}; see Appendix~\ref{app:related} for extended discussion of prior work.

\section{Methods}
\label{sec:setup}

\subsection{Decoding Metrics}
\label{sec:metrics}

We evaluate eight complementary metrics organized into two groups: four \textit{static} metrics that operate on time-averaged representations and four \textit{trajectory} metrics that characterize temporal dynamics: k-NN accuracy, RSA, CKA, Procrustes R$^2$, time-resolved RSA, Speed Profile Correlation, Trajectory Procrustes R$^2$, and Dynamical Similarity Analysis. All metrics are normalized or correlation-based and therefore do not confound with subpopulation size, and constructed such that high values mean higher alignment. Definitions and implementation details are given in Appendix~\ref{app:dec_metrics}.

\begin{figure*}[t]
    \centering
    \includegraphics[width=\linewidth]{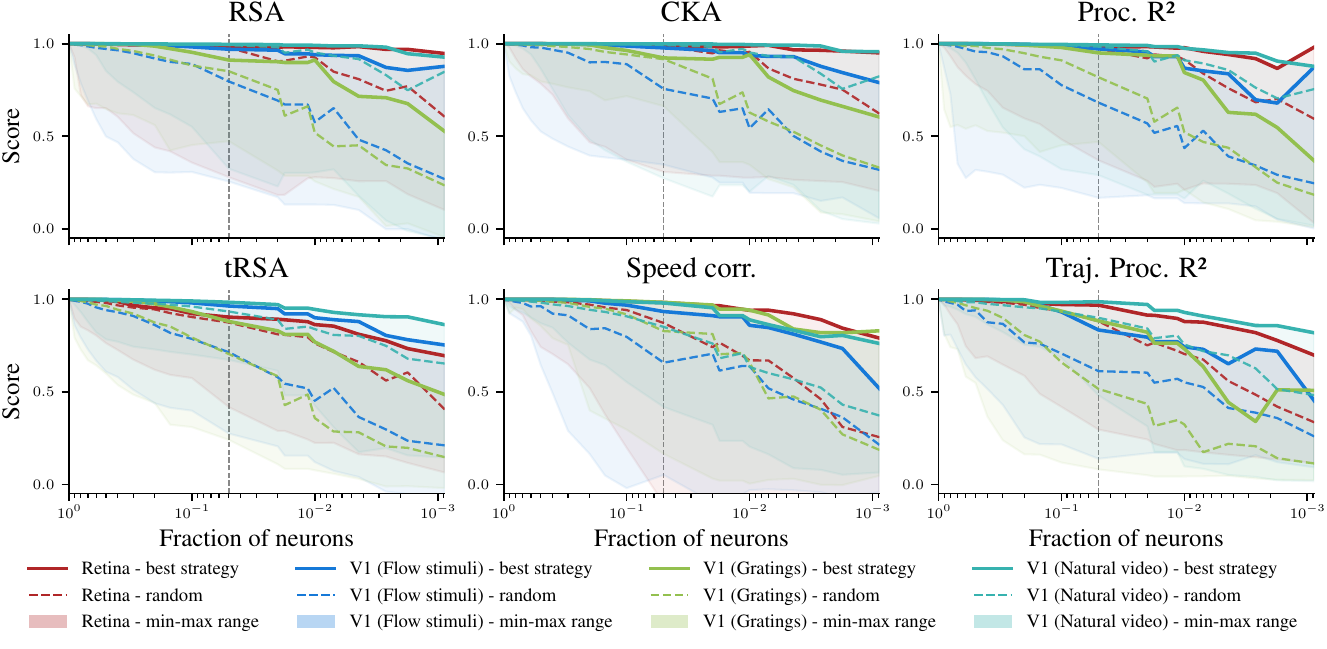}
    \caption{Population-size sweep across Retina, V1 (flows, gratings, natural) for six decoding metrics. Solid lines: best-performing selection strategy; dashed lines: random selection; shaded bands: min--max range across all strategies. The vertical dashed line marks 5\% of neurons (used in Figure~\ref{fig:regions5}). Static metrics plateau near ceiling at 5\% or below under the best strategy.}
    \label{fig:sweep}
\end{figure*}

\subsection{Encoding Manifold}
\label{sec:encoding_manifold}

The \textit{encoding manifold} \citep{dyballaPopulationEncodingStimulus2024, bertramHowNeuralNeural2026} describes how neurons are organized in \textit{neural response space} rather than stimulus space. The encoding manifold is the point cloud $\mathcal{E}$; the global topology of $\mathcal{E}$ --- in particular, whether neurons cluster into distinct functional groups and how those groups are arranged --- captures the internal organization of the population independently of its aggregate decoding performance. We construct it via nonnegative tensor factorization $\to$ IAN adaptive graph \citep{dyballaIANIteratedAdaptive2023} $\to$ diffusion maps \citep{coifman2005geometric, coifmanDiffusionMaps2006}, following \citet{dyballaPopulationEncodingStimulus2024}; full details are given in Appendix~\ref{app:encoding_pipeline}. The encoding manifold is biologically grounded: applied to retina it recovers the known retinal ganglion cell types as distinct clusters, whereas mouse V1 --- which has no known discrete functional types --- forms a smooth continuum \citep{dyballaPopulationEncodingStimulus2024, nellenLearningClusterNeuronal2025}. This validates it as a descriptor of functional organization rather than an arbitrary embedding.

We measure encoding manifold similarity via Gromov--Wasserstein (GW) distance \citep{peyreComputationalOptimalTransport2020, memoliGromovWassersteinDistances2011}, applied here as an intrinsic measure of encoding manifold coverage rather than cross-system alignment. GW measures how well the internal pairwise-distance structure of one point cloud can be matched to another: two manifolds with the same shape score zero, while structurally incompatible ones score high. Normalized GW similarity is:

$$
\mathrm{GW}(\mathcal{X},\mathcal{Y})
= \sqrt{\inf_{\pi\in\Pi(\mu,\nu)}\!
  \sum_{i,i',j,j'} |\mathbf{C}_\mathcal{X}^{ii'} - \mathbf{C}_\mathcal{Y}^{jj'}|^2
  \,\pi_{ij}\pi_{i'j'}},
$$
$$
\mathrm{GW_{Sim}}(\mathcal{E}_\mathrm{sub})
= 1-\frac{\mathrm{GW}(\mathcal{E}_\mathrm{sub},\,\mathcal{E}_\mathrm{full})}
  {\mathrm{GW}_\mathrm{max}},
$$

where $\mathcal{E}_\mathrm{sub}$ and $\mathcal{E}_\mathrm{full}$ are encoding manifolds of the subpopulation and full population, and $\mathrm{GW}_\mathrm{max} = \mathrm{GW}(\mathcal{E}_\mathrm{full},\,\mathcal{E}_\mathrm{coll})$ normalizes by a collapsed-centroid baseline ($\epsilon = 10^{-8}$ jitter to prevent degeneracy).

\section{Results}
\label{sec:results}

\subsection{Causal Dissociation via Controlled Training}
\label{sec:mnist}

\begin{figure*}[t]
    \centering
    \includegraphics[width=\linewidth]{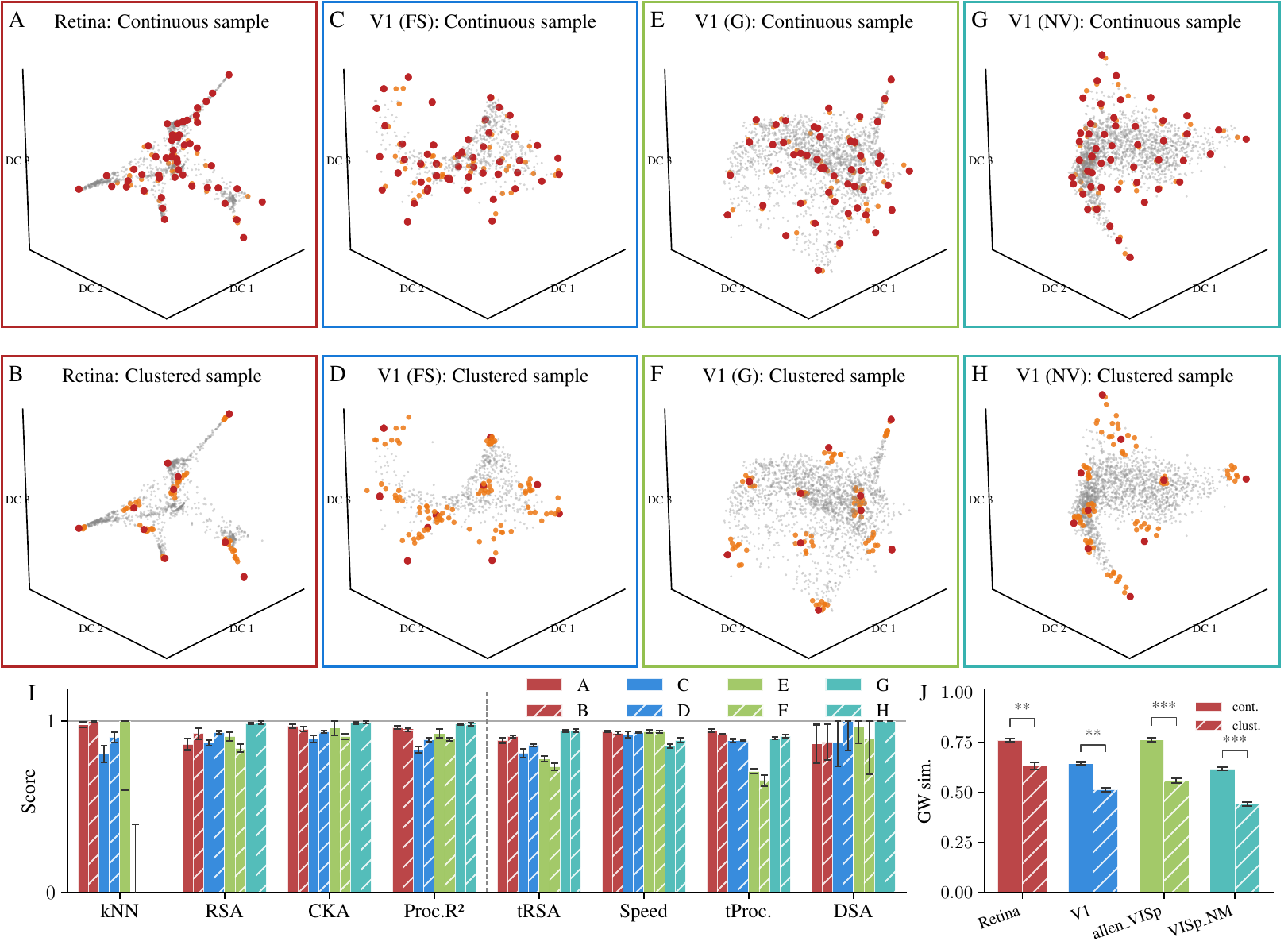}
    \caption{FPS-based subpopulations with continuous (top row, dark red) vs.\ clustered (bottom row, orange) encoding manifold topology for Retina, V1, VISp (gratings), and VISp (natural). Continuous subpopulations are sparsely distributed across the full manifold; clustered subpopulations are concentrated in local neighborhoods. Bar chart shows all eight metrics for both conditions; error bars are 1 s.d.\ over FPS seeds. Scores are near-ceiling for both topologies across nearly all systems and metrics, demonstrating that decoding metrics are insensitive to this fundamental difference in encoding structure. (Wilcoxon rank-sum; ** $p{<}0.01$, *** $p{<}0.001$)}
    \label{fig:fps}
    \vspace{-2mm}
\end{figure*}

We train a three-layer CNN on MNIST digit classification (10 classes, $\approx 60\,000$ images; details in Appendix~\ref{app:mnist_method}) and add an auxiliary loss $\mathcal{L}_\mathrm{cluster}(\lambda)$ that progressively pulls hidden-layer neurons toward digit-specialist response templates, systematically clustering the encoding manifold \citep{khoslaSupervisedContrastiveLearning2021}.

$$
\mathcal{L}_\text{cluster} = \frac{1}{|\mathcal{A}|} \sum_{i \in \mathcal{A}} \frac{1}{|P(i)|} \sum_{p \in P(i)} -\log \frac{\exp(\mathbf{z}_i \cdot \mathbf{z}_p / \tau)}{\sum_{j \neq i} \exp(\mathbf{z}_i \cdot \mathbf{z}_j / \tau)},
$$

where $\mathbf{z}_i = \mathbf{h}_i / \|\mathbf{h}_i\|$ is the $\ell_2$-normalized activation profile of neuron $i$ (a $B$-dimensional vector), $P(i) = \{j \neq i : \text{pref}(j) = \text{pref}(i)\}$ is the set of neurons sharing neuron $i$'s preferred digit, $\mathcal{A} = \{i : |P(i)| > 0\}$ is the set of anchor neurons with at least one positive, and $\tau = 0.07$ is the temperature. Task accuracy is held approximately constant ($\approx 98.5\%$) across all values of $\lambda$.

Table~\ref{tab:mnist_main} reports decoding metrics and GW encoding similarity as $\lambda$ increases from 0 to 50. GW similarity drops monotonically from 1.00 to 0.28 --- the previously continuous population response space becomes sharply clustered around digit-specialist prototypes (Figure~\ref{fig:mnist}) --- while RSA, CKA, and Procrustes $R^2$ remain above 0.93 up to $\lambda{=}10$. We focus on $\lambda{=}10$ vs.\ baseline (20 seeds each) for the main analysis.

\textbf{Functional Differences.} Despite matched decoding behavior, the two models differ functionally. Neurons in the clustered model show significantly lower mean tuning and higher variance (Figure~\ref{fig:mnist} C, D), reflecting sharper, class-selective responses (B) vs.\ graded, multi-class tuning (A). Gradient-activation attribution (Table~\ref{tab:mnist_main}; Appendix~\ref{app:mnist_attribution}) reveals that the constrained model uses more class-specific circuits: attribution entropy and overlap are significantly lower ($p{<}0.001$) and weight-graph assortativity significantly higher ($p{<}0.001$), indicating same-class neurons preferentially connect within class-specific pathways. Out-of-distribution blur robustness also differs significantly ($p{<}0.05$). Together, two models indistinguishable by RSA, CKA, or task accuracy implement their shared behavior through fundamentally different internal architectures.

\subsection{Decoding Metrics Saturate with Small Subpopulations}
\label{sec:sweep}

We next show that decoding scores can be reproduced by a small fraction of neurons. We perform a population-size sweep under six neuron-selection strategies (Appendix~\ref{app:sweep_method}). Across biological systems (Figure~\ref{fig:sweep}), all metrics saturate near ceiling at or above 5\% of neurons under information-maximizing strategies; RSA and CKA remain near ceiling even under random selection. This motivates asking what the remaining neurons implement --- the following experiment tests this directly.

\subsection{Encoding Manifold Topology Can Be Disrupted While Preserving Decoding Metrics}
\label{sec:fps}

\looseness=-1
We use \textit{Farthest Point Sampling} (FPS) to construct subpopulations (Appendix~\ref{app:fps_method}): by varying $n$ seed neurons (breadth of manifold coverage) and $m$ nearest neighbors per seed, we interpolate between a \textbf{continuous} sample that spans the full encoding manifold ($m{=}1$) and a locally \textbf{clustered} sample ($m{>}1$). Figure~\ref{fig:fps} shows results for Retina, V1, and Allen V1 (VISp).

The correlation-based decoding metrics remain near-ceiling for both functional topologies across systems (Figure~\ref{fig:fps}). Together, Retina, V1, and ten Allen cortical datasets (five areas $\times$ two stimulus classes) establish that the insensitivity of RSA, CKA, DSA and other decoding metrics to encoding manifold topology is robust and general. The one exception is k-NN accuracy, a hard classifier rather than a correlation: it can drop sharply for clustered samples (e.g.\ VISp gratings, $0.80 \to 0.20$ in Table~\ref{tab:showcase}). Machine learning models including FNN, Flyvision, and R(2+1)D show the same pattern (Appendix~\ref{app:ml_fps}). Full per-dataset scores and manifold visualizations for all 17 datasets are given in Table~\ref{tab:showcase} and Figures~\ref{fig:retina_fps}--\ref{fig:r2plus1d_fps}.

Where all these decoding metrics fail to differentiate, our GW similarity provides a significant distinction between continuous and clustered samples in all settings (Figure~\ref{fig:fps} J). The continuous sample always spans the full encoding manifold better and is thus measurably more similar in function to the full population. GW similarity thus provides a complementary, encoding-side measure that captures precisely what decoding metrics are blind to. The region-sampling experiment (Appendix~\ref{app:regions}) further shows that encoding manifold position predicts decoding fidelity, ruling out redundancy as an explanation.

We provide a \href{https://johannesbertram.github.io/FNN_Manifolds/index.html}{Neural Manifold Explorer} tool for further experimentation: an interactive 3-D viewer for inspecting the encoding and decoding manifolds and subpopulations of every system analyzed here.

\subsection{Vision Transformers}

\begin{figure}
    \centering
    \includegraphics[width=\linewidth]{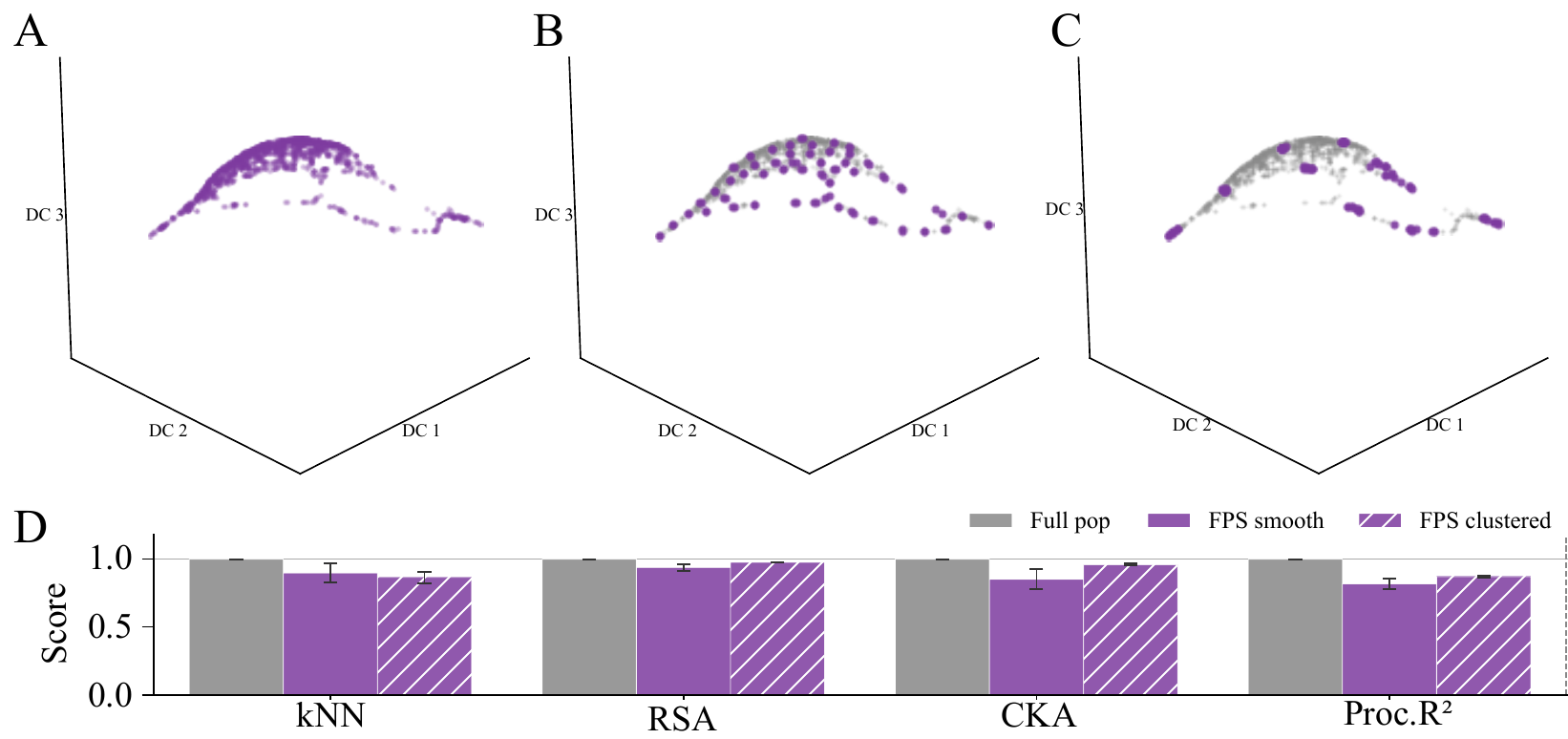}
    \caption{FPS-based subpopulations with continuous (B) and clustered (C) encoding manifold topology for a ViT. Stationary decoding metrics are high across conditions and indistinguishable between the two encoding manifold topologies.}
    \label{fig:vit_main}
\end{figure}

\looseness=-1
Our pipeline assumes a $\text{neuron} \times \text{stimulus} \times \text{direction} \times \text{time}$ response tensor, but a vision transformer (ViT) \citep{dosovitskiyImageWorth16x162021} has neither a canonical neuron axis nor a time axis. We obtain our data tensor from relating hidden dimensions and tokens in the ViT to channels and within-channel neurons in CNNs: We map a ViT onto our framework by treating each (token, hidden-dimension) entry as a ``neuron'' (up to $197 \times 768$ candidate units, of which 2000 are retained by max-activation sampling), and by supplying the temporal axis either from feeding frame sequences through a fixed encoder block or from treating successive encoder layers as time (the latter ``layers-as-time'' framing is detailed for the Raptor model; Appendix~\ref{app:data}, Figures~\ref{fig:vit},~\ref{fig:raptor}). With this mapping, the same pattern reproduces (Figure~\ref{fig:vit_main}, Table~\ref{tab:showcase}): correlation-based decoding metrics remain high and statistically indistinguishable between FPS-continuous and FPS-clustered encoding topologies. We make no claim here about the specific structure of ViT encoding manifolds; rather, this demonstrates that the encoding-manifold diagnostic is architecture-agnostic and can be applied to transformer models.

\vspace{-1mm}
%------------------------------------------------------------
\section{Discussion}
\label{sec:discussion}
%------------------------------------------------------------
\vspace{-1mm}

\textbf{Implications for model--brain comparisons: behavioral \textit{vs.} functional alignment.} Current evaluations of neural models using RSA, CKA, or decoding accuracy establish \emph{behavioral} alignment---what the population does, collectively---but not \emph{functional} alignment---how that behavior is implemented. A model may match biological responses while implementing a fundamentally different internal organization: distinct functional subpopulations, altered inter-group relationships, and different representational roles distributed across neurons. The MNIST experiment makes this concrete: models matched on decoding metrics differ significantly in tuning sharpness, out-of-distribution robustness, and the structure of their computational circuits (Section~\ref{sec:mnist}). We therefore recommend complementing decoding-based comparisons with encoding manifolds and GW-based measures of functional organization.

\looseness=-1
\textbf{Implications for mechanistic interpretability.} Circuit-tracing methods ask which neurons and pathways implement specific behaviors \citep{Ameisenetal2025, Lindseyetal2024, dunefskyTranscodersFindInterpretable2024}. Our results show that standard alignment metrics are blind to precisely what circuit analysis seeks to reveal: the internal functional organization of a population. Our MNIST experiment makes this concrete: models statistically indistinguishable by RSA and CKA differ significantly in tuning, attribution structure, and weight-graph assortativity. We do not claim the encoding manifold identifies causal circuits; rather, it provides a scalable, geometric complement --- by organizing neurons by response profiles rather than by stimuli, it surfaces differences in functional organization that decoding metrics miss. We propose encoding manifolds and GW-based similarity as scalable complements to circuit tracing for assessing functional equivalence, not merely behavioral equivalence.

\looseness=-1
\textbf{Conclusion.} Our results do not invalidate decoding-based similarity metrics; rather, they delimit what those metrics can support. RSA, CKA, and related measures are informative about how well a population can be read out under a given stimulus set, but they do not, by themselves, determine how that population is internally organized. Encoding manifolds provide a complementary description of that organization, and GW gives a principled way to compare it across systems. In this sense, the appropriate unit of comparison is not a single scalar alignment score, but a pair of measurements: what a system does, and how it is organized to do it.

\section{Acknowledgments and Disclosure of Funding}
%Use unnumbered first level headings for the acknowledgments. All acknowledgments go at the end of the paper before the list of references. Moreover, you are required to declare funding (financial activities supporting the submitted work) and competing interests (related financial activities outside the submitted work). More information about this disclosure can be found at: \url{https://neurips.cc/Conferences/2025/PaperInformation/FundingDisclosure}.

%Do {\bf not} include this section in the anonymized submission, only in the final paper. You can use the \texttt{ack} environment provided in the style file to automatically hide this section in the anonymized submission.

This article has received funding from the European Commission's Marie Sk\l{}odowska-Curie Action under grant agreement no. 101207931 (LD). SWZ was supported by NIH grant 1R01EY031059 and NSF Grant 1822598.

\bibliography{bib}
\bibliographystyle{icml2026}

%%%%%%%%%%%%%%%%%%%%%%%%%%%%%%%%%%%%%%%%%%%%%%%%%%%%%%%%%%%%
\newpage
\appendix

\section{Extended Related Work}
\label{app:related}

RSA \citep{kriegeskorteRepresentationalSimilarityAnalysis2008} and CKA \citep{kornblithSimilarityNeuralNetwork2019} have become central tools for the comparison of neural representation, distilling high-dimensional population responses into tractable similarity measures that facilitate large-scale cross-system analyses. However, by construction, they operate on population-level summaries of activity, potentially obscuring the internal structure and functional heterogeneity of the underlying units. This is reflected in a growing body of work questioning their statistical and theoretical properties: \citet{dingGroundingRepresentationSimilarity2021, ahlertHowAlignedAre2024, boEvaluatingRepresentationalSimilarity2025, hofling2026only} show that common measures frequently disagree, while \citet{murphyCorrectingBiasedCentered2024} identify a bias that drives CKA toward its maximum in the low-sample, high-dimensional regime typical of neural data. On the theoretical side, \citet{harveyWhatRepresentationalSimilarity2024} show that CKA, CCA, and Procrustes all quantify alignment of optimal linear readouts, exposing their decoding-centric nature, and \citet{almudevarBridgingFunctionalRepresentational2026} unify them under a usable-information framework. \citet{avitanModelBehaviorAlignmentFlexible2025} show in model-recovery experiments that the highest-scoring model by linear readout is frequently not the correct generative model, and \citet{featherBrainModelEvaluationsNeed2025} propose requiring models to match internal neural representations within individual variability margins rather than aggregate decoding scores. Related work \citet{davariReliabilityCKASimilarity2022, harveyWhatRepresentationalSimilarity2024, boEvaluatingRepresentationalSimilarity2025} has proposed comparing systems via alignment of their output functions. In the interpretability literature, circuit tracing asks a similar question by investigating the function implemented by a neural network \citep{Ameisenetal2025, Lindseyetal2024, dunefskyTranscodersFindInterpretable2024}. We emphasize that the decoding-centric --- or linear-readout --- nature of RSA, CKA, and Procrustes is thus already established in this prior work, and that these metrics were never intended to assert mechanistic or circuit equivalence. Our contribution is therefore not this observation itself, but two constructive additions: (i) a controlled, \emph{causal} dissociation in which encoding topology is manipulated by design while decoding metrics remain matched (Section~\ref{sec:mnist}), and (ii) an encoding-side measure --- the encoding manifold compared via Gromov--Wasserstein distance --- that captures the internal organization to which these metrics are blind. Our work is thus complementary: we shift the focus from output alignment and task performance to internal organization, analyzing functional topology within neural populations to reveal differences not captured by traditional similarity metrics.

\section{Limitations}

First, the MNIST experiment uses a specific clustering loss and a relatively shallow architecture; whether the same decoupling holds for deeper networks or other training objectives remains open. Nevertheless, it provides a concrete existence proof that encoding and decoding metrics can be dissociated, motivating further investigation. The clustering loss also alters tuning sharpness and inter-neuron connectivity simultaneously; an ablation that isolates encoding-manifold topology from the regularizer's direct effect on weight magnitudes (e.g.\ matching tuning sharpness via L1 or dropout) is left to future work. Furthermore, GW is used here as an intrinsic measure of encoding-manifold coverage and has not yet been validated against ground-truth circuit differences in a cross-system, model--brain comparison, which we leave to future work. Finally, although GW provides a practical measure of encoding similarity, its behavior in more diverse settings remains to be validated in future work.

\section{Additional Results}
\label{app:additional_results}

\begin{figure*}
    \centering
    \includegraphics[width=\linewidth]{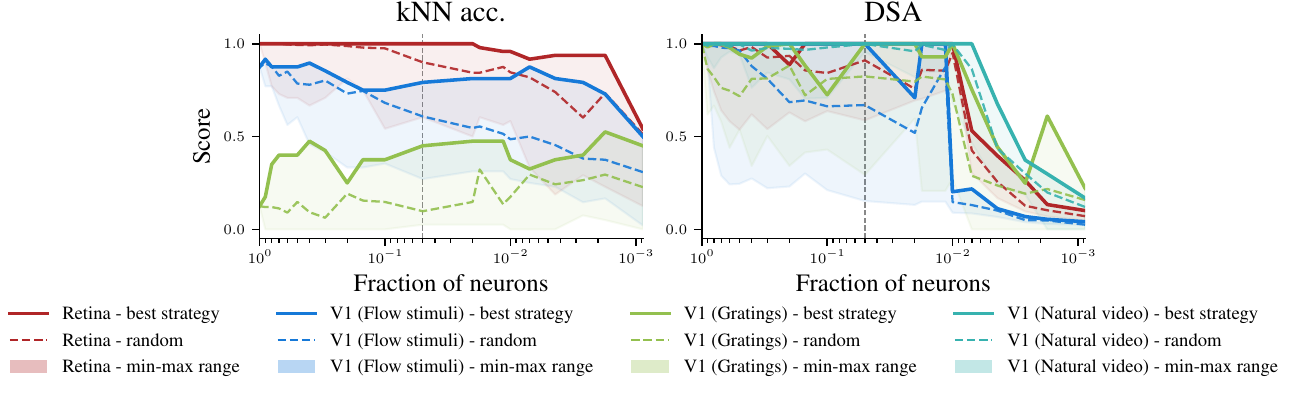}
    \caption{Fraction sweep for kNN accuracy and normalized DSA z scores.}
    \label{app:sweep}
\end{figure*}

\begin{figure*}
    \centering
    \includegraphics[width=0.75\linewidth]{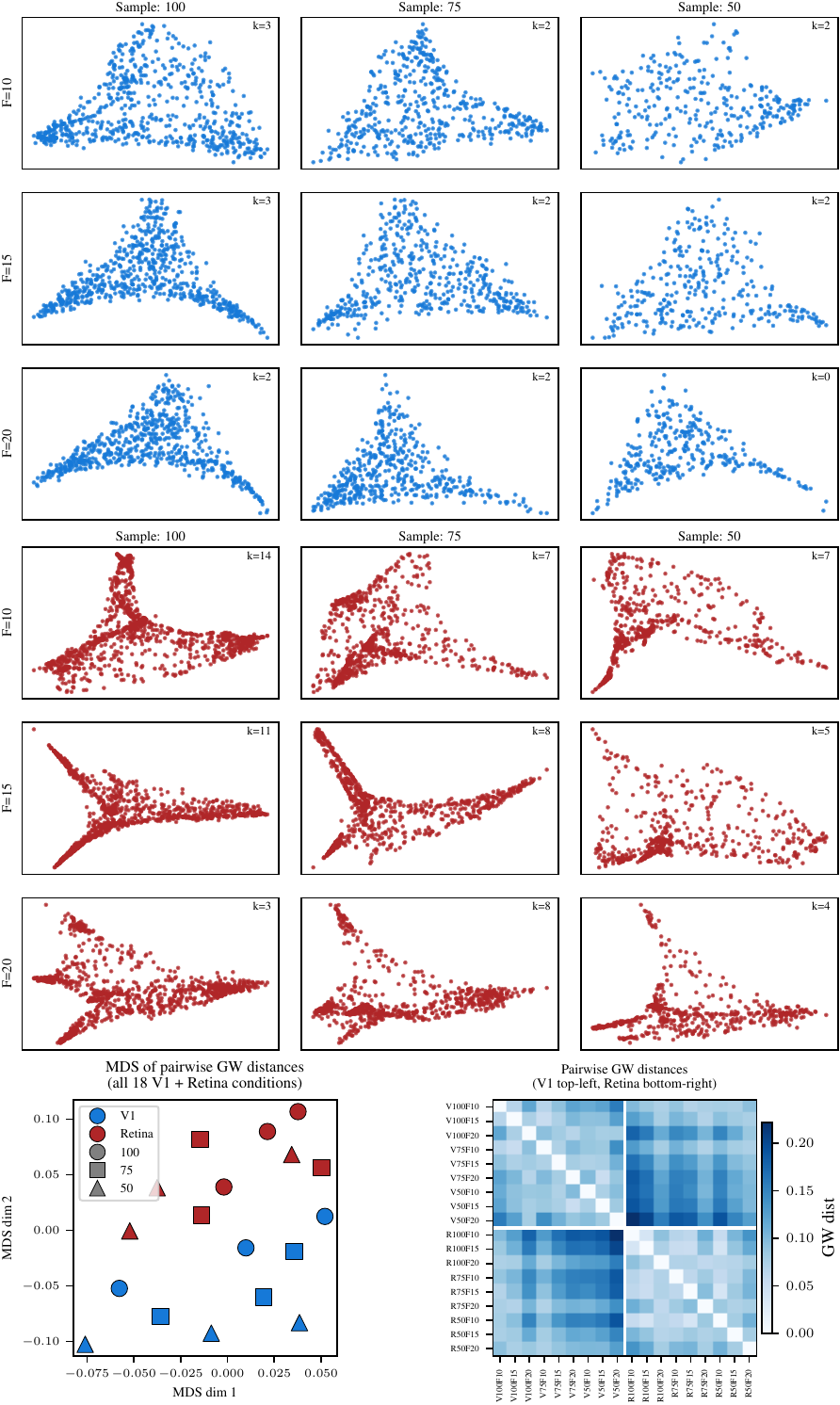}
    \caption{\textbf{Encoding manifold pipeline stability} across sample size (columns), tensor decomposition factors (rows) and datasets (Blue: V1, Red: Retina). Encoding topology is stable, showing continuous manifolds in V1 and clustered ones in retina. GW distance groups the two datasets together (bottom).}
    \label{fig:enc_stability}
\end{figure*}

\begin{figure*}
    \centering
    \includegraphics[width=\linewidth]{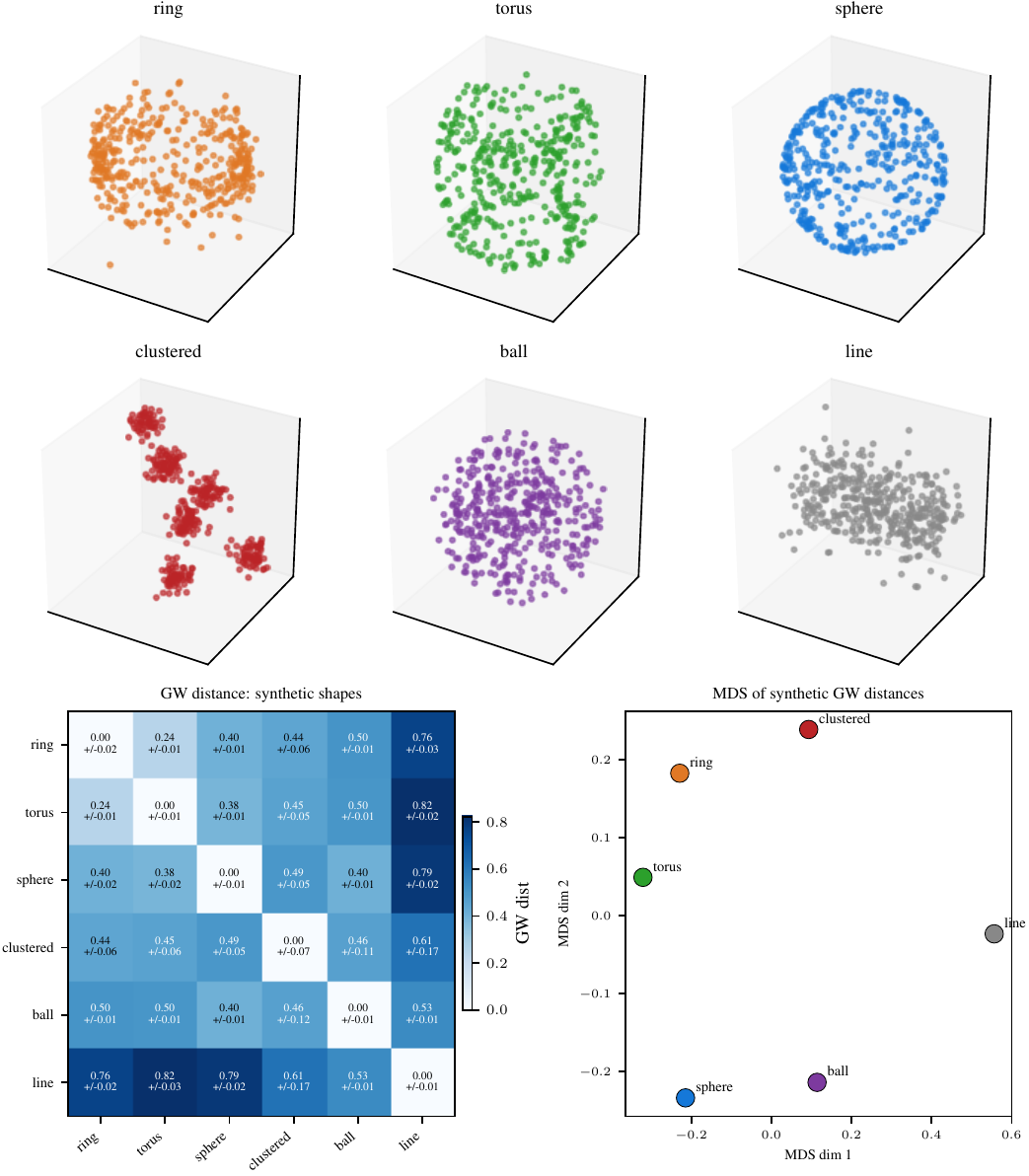}
    \caption{\textbf{GW intuition} based on simple synthetic datasets. The ring and torus being topologically equivalent are grouped together. Similarly, the sphere and ball are close in GW space. The line is furthest from all other datasets, not allowing for circular matching of points.}
    \label{fig:gw_int}
\end{figure*}

\subsection{Encoding Manifold Position Determines Decoding Fidelity}
\label{app:regions}

\begin{figure*}
    \centering
    \includegraphics[width=\linewidth]{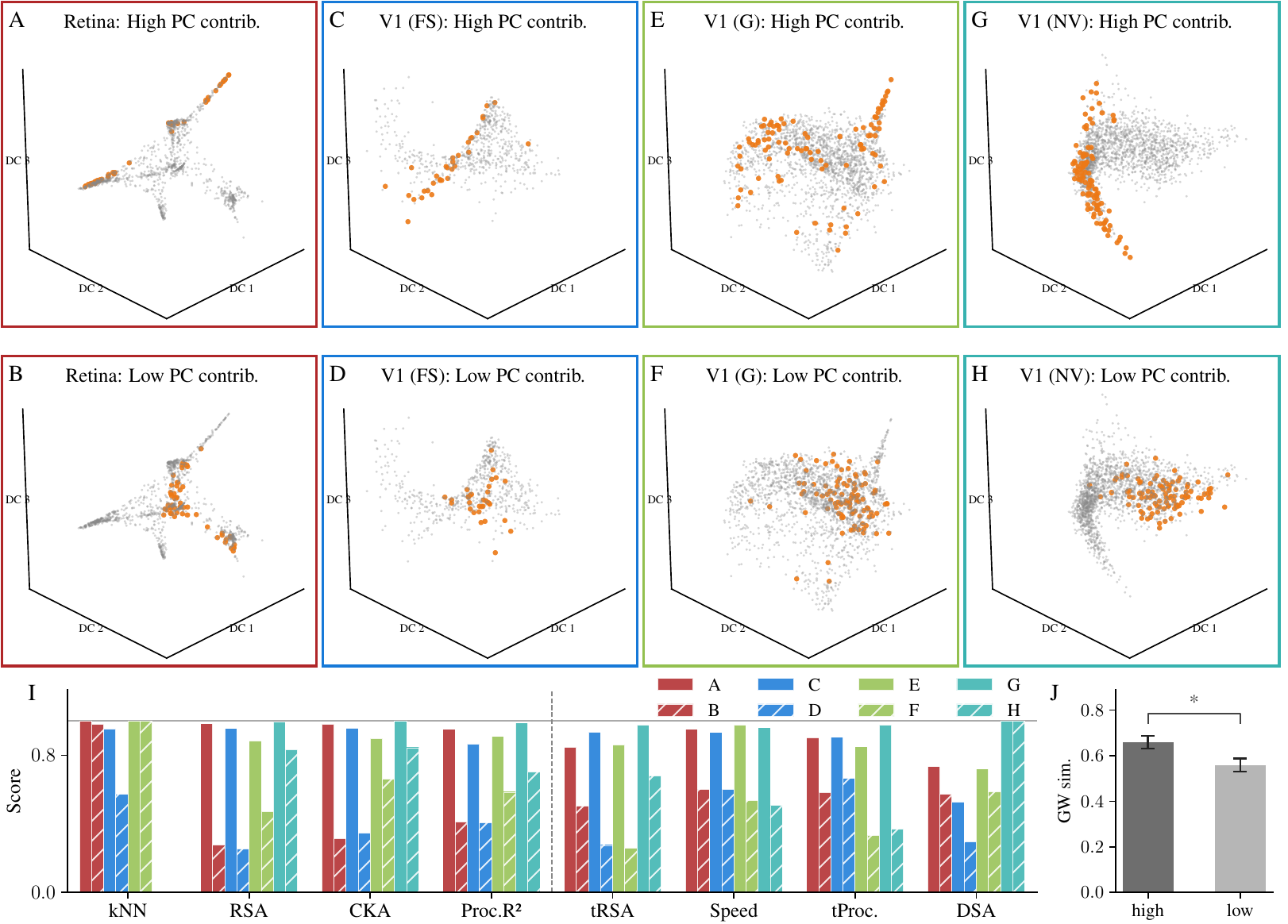}
    \caption{
    \looseness=-1
    Encoding manifold coverage and decoding scores for 5\% subpopulations selected by highest (top row, solid bars) or lowest (bottom row, hatched bars) first-PC contribution in decoding space, for Retina, V1, VISp (gratings), and VISp (natural). Orange dots show selected neurons on the encoding manifold (gray: full population); in all cases the selection covers a small, localized region of the manifold. Bar chart shows all eight metrics; RSA and CKA show the largest differences between conditions. Robustness across sizes in Figures~\ref{fig:regions1} (1\%) and~\ref{fig:regions10} (10\%) (Wilcoxon rank-sum; * $p{<}0.05$).}
    \label{fig:regions5}
    \vspace{-2mm}
\end{figure*}

To link subpopulation decoding fidelity to encoding manifold coverage, in Figure \ref{fig:regions5}, we select size-matched subpopulations by maximizing (top row, A/C/E/G) or minimizing (bottom row, B/D/F/H) each neuron's contribution to the first principal component of the decoding space --- yielding near best- and worst-case samples for decoding metrics, respectively (Appendix~\ref{app:region_method}).

\begin{figure*}
    \centering
    \includegraphics[width=\linewidth]{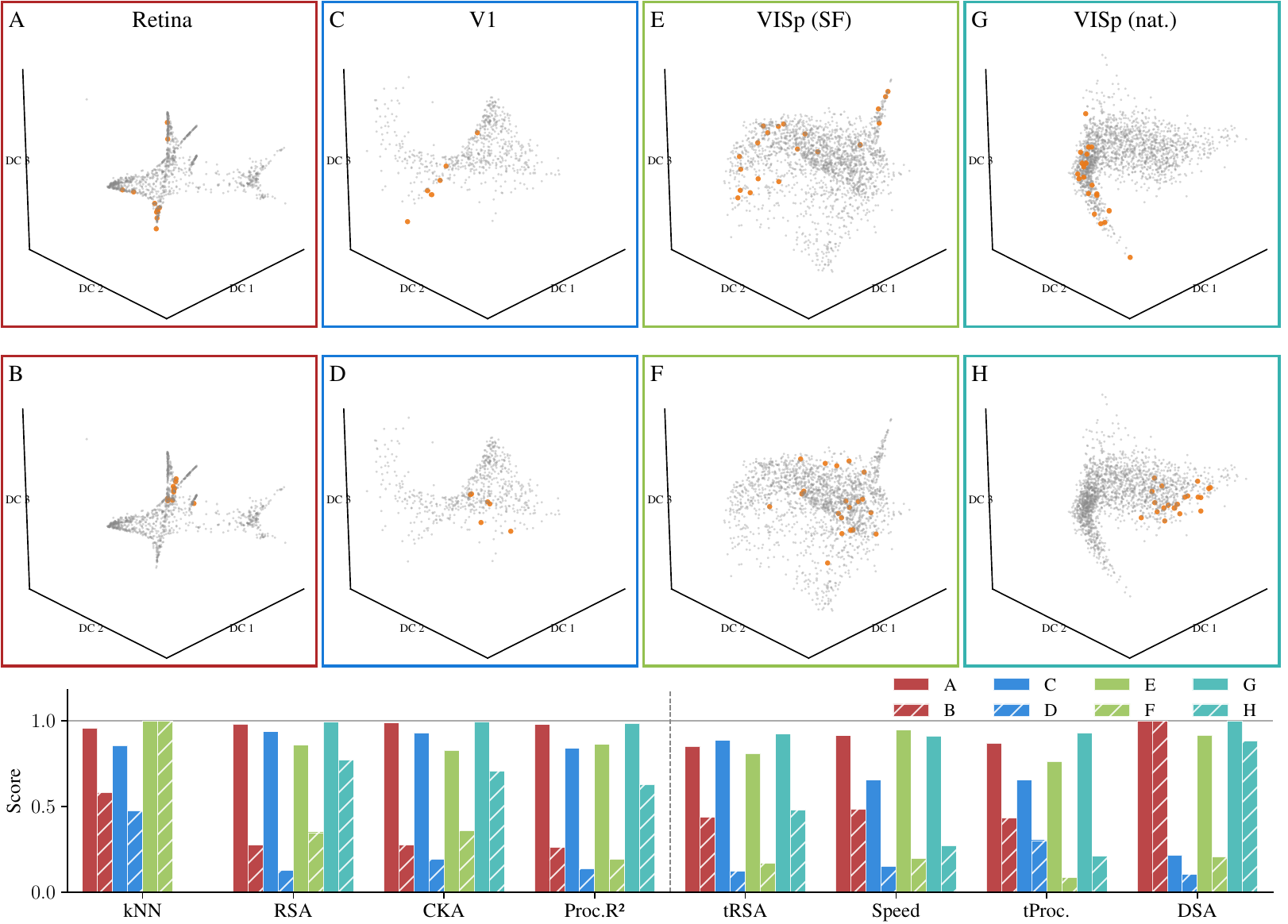}
    \caption{Size-matched encoding manifold region sampling for Retina, V1 and VISp (gratings and natural video). Samples drawn to maximize (top) or minimize (bottom) first principal component contribution of selected neurons, generating near best-/worst-case samples. These samples from distinct regions of the encoding manifold yield qualitatively different decoding profiles under the same eight metrics. In all conditions, a small (1\%) subregion sample can suffice to largely recover alignment metrics.}
    \label{fig:regions1}
\end{figure*}

\begin{figure*}
    \centering
    \includegraphics[width=\linewidth]{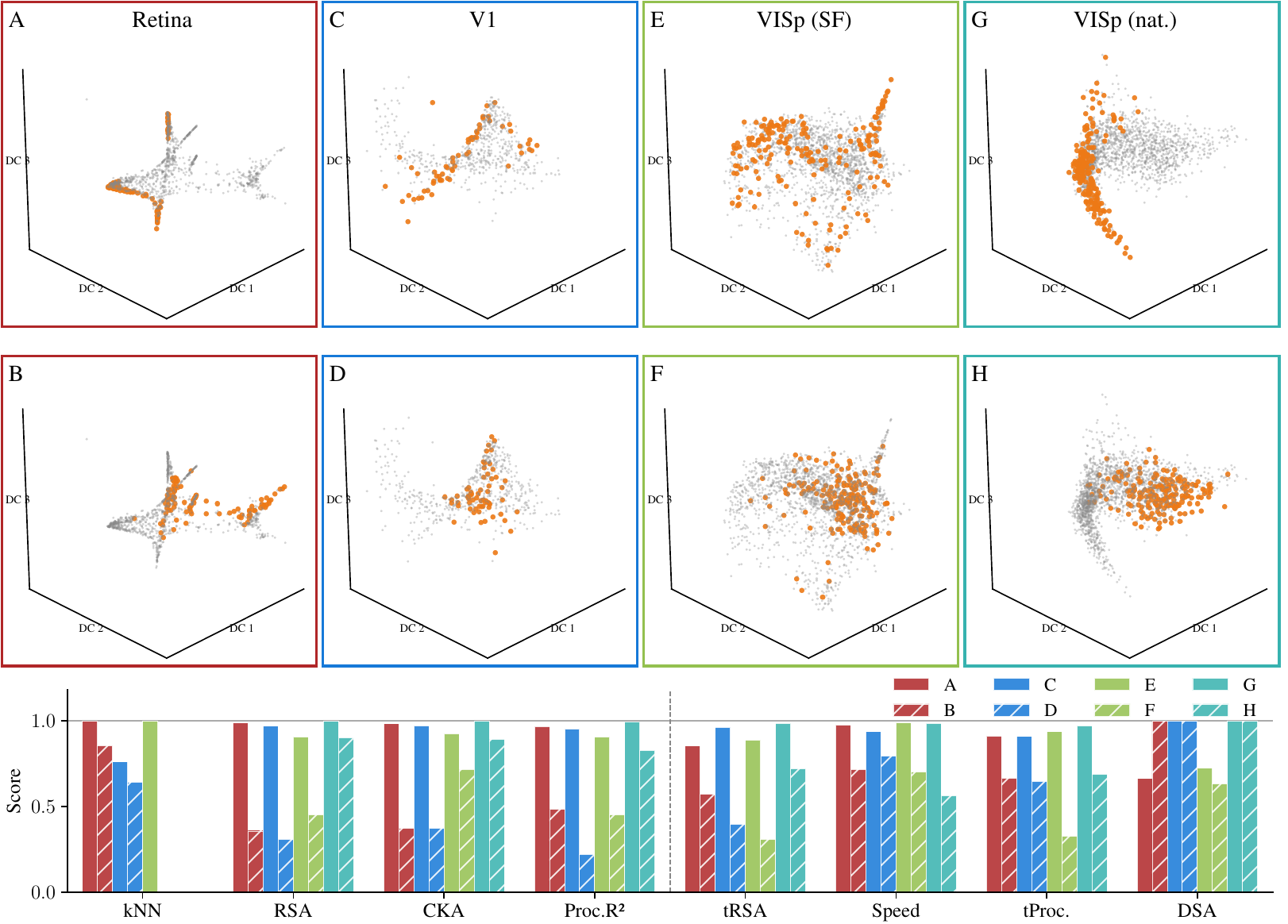}
    \caption{Size-matched encoding manifold region sampling for Retina, V1 and VISp (gratings and natural video). Samples drawn to maximize (top) or minimize (bottom) first principal component contribution of selected neurons, generating near best-/worst-case samples. These samples from distinct regions of the encoding manifold yield qualitatively different decoding profiles under the same eight metrics. In all conditions, a small (10\%) subregion sample can suffice to largely recover alignment metrics.}
    \label{fig:regions10}
\end{figure*}

The bar chart confirms this selection produces a wide range of decoding scores: high-PC1 subpopulations largely recover the full-population profile across all eight metrics, while low-PC1 subpopulations yield substantially lower scores. The encoding manifold panels tell the complementary story: regardless of whether a subpopulation scores well on decoding metrics, it occupies only a small, localized region of the encoding manifold. This holds for all four systems and both stimulus classes (see Figures~\ref{fig:regions1} and~\ref{fig:regions10} for 1\% and 10\% fractions). In this setting where decoding metrics can capture differences between the high-PC1 and low-PC1 conditions, GW can also capture this effect. The high-PC1 subpopulations are significantly closer to the full population as measured by GW self similarity of each subset of neurons to the full set (Figure~\ref{fig:regions5} J). However, this difference is less than in many of the decoding metrics, backing up the claim that these metrics are complimentary.

\looseness=-1
\textbf{Encoding manifold topology characterizes function.} However, the encoding manifold, by design, organizes the full population of neurons by function, and has been validated against known results from biology. For example, the retinal encoding manifold clusters neurons by known retinal ganglion cell types \citep{dyballaPopulationEncodingStimulus2024}. Since there are no known functional groups of neurons in mouse V1 \citep{dyballaPopulationEncodingStimulus2024, nellenLearningClusterNeuronal2025}, the encoding manifold continuously organizes V1 neurons. These stable biological properties of computation are robustly captured by the encoding manifold pipeline (Appendix~\ref{app:enc_robustness}, Figure~\ref{fig:enc_stability}). Therefore, we argue that encoding manifold coverage and topology are relevant properties for functional alignment. Our artificial MNIST example provides additional evidence for these claims in a controlled setting (Section \ref{sec:mnist}). As a result, sampling region-constrained subpopulations on the encoding manifold omits biologically relevant functions, such as throwing away the majority of retinal ganglion cell types. Still, these functionally misaligned populations yield behaviorally aligned representations, showing the dissociation between functional and behavioral alignment.

\begin{table*}[h]
\centering
\small
\caption{MNIST $\lambda$-sweep: encoding topology is progressively clustered while decoding metrics remain near-ceiling. All scores are relative to the baseline ($\lambda=0$). GW Sim (IAN) measures normalized encoding manifold similarity via Gromov--Wasserstein distance.}
\label{tab:mnist_sweep}
\begin{tabular}{r r r r r r}
\toprule
$\lambda$ & Acc & RSA & CKA & Proc.\ $R^2$ & GW Sim (IAN) \\
\midrule
0 (baseline) & 0.985 & 1.000 & 1.000 & 1.000 & 1.000 \\
0.1          & 0.986 & 0.997 & 0.998 & 0.997 & 0.970 \\
0.5          & 0.986 & 0.997 & 0.998 & 0.998 & 0.941 \\
1            & 0.985 & 0.996 & 0.998 & 0.997 & 0.943 \\
5            & 0.985 & 0.982 & 0.990 & 0.989 & 0.706 \\
10           & 0.984 & 0.968 & 0.982 & 0.981 & 0.466 \\
50           & 0.983 & 0.897 & 0.936 & 0.925 & 0.284 \\
\bottomrule
\end{tabular}
\end{table*}

\begin{table*}[h]
\centering
\small
\caption{Gaussian blur robustness: test accuracy at each blur level for baseline ($\lambda{=}0$) vs.\ clustered ($\lambda{=}10$), $n{=}20$ seeds (mean\,$\pm$\,SEM, Wilcoxon paired; ns\,$=$\,$p{\geq}0.05$).}
\label{tab:blur}
\begin{tabular}{r r r r}
\toprule
$\sigma_{\mathrm{blur}}$ & Baseline & Clustered ($\lambda{=}10$) & $p$ \\
\midrule
0 & $0.986 \pm 0.000$ & $0.983 \pm 0.000$ & $0.004$ ** \\
2 & $0.870 \pm 0.011$ & $0.873 \pm 0.016$ & $0.846$ ns \\
3 & $0.621 \pm 0.021$ & $0.610 \pm 0.026$ & $0.625$ ns \\
4 & $0.432 \pm 0.017$ & $0.387 \pm 0.022$ & $0.160$ ns \\
5 & $0.305 \pm 0.017$ & $0.256 \pm 0.021$ & $0.064$ ns \\
6 & $0.216 \pm 0.017$ & $0.204 \pm 0.020$ & $0.695$ ns \\
\bottomrule
\end{tabular}
\end{table*}

\begin{table*}[htbp]
\centering
\small
\caption{Subpopulation metric scores across 17 datasets. Top rows (continuous): FPS 50 seeds $\times$ 1\,NN; Bottom rows (Clustered): 10 seeds $\times$ 9\,NN. Values show mean $\pm$ std over 5 seeds.}
\label{tab:showcase}
\begin{tabular}{lrrrrrrrr}
\toprule
Dataset & kNN & RSA & CKA & Proc.$R^2$ & tRSA & Spd. & tProc & DSA \\
\midrule
VISp G& 0.80{\scriptsize$\pm$.40} & 0.88{\scriptsize$\pm$.03} & 0.92{\scriptsize$\pm$.04} & 0.90{\scriptsize$\pm$.02} & 0.76{\scriptsize$\pm$.02} & 0.94{\scriptsize$\pm$.01} & 0.69{\scriptsize$\pm$.01} & 0.88{\scriptsize$\pm$.10} \\
Cluster & 0.20{\scriptsize$\pm$.40} & 0.88{\scriptsize$\pm$.02} & 0.93{\scriptsize$\pm$.01} & 0.92{\scriptsize$\pm$.01} & 0.75{\scriptsize$\pm$.02} & 0.94{\scriptsize$\pm$.01} & 0.68{\scriptsize$\pm$.03} & 0.74{\scriptsize$\pm$.20} \\
\midrule
VISrl G& 1.00 & 0.78{\scriptsize$\pm$.03} & 0.89{\scriptsize$\pm$.03} & 0.87{\scriptsize$\pm$.02} & 0.65{\scriptsize$\pm$.02} & 0.94{\scriptsize$\pm$.01} & 0.62{\scriptsize$\pm$.01} & 0.42{\scriptsize$\pm$.20} \\
 Cluster & 1.00 & 0.77{\scriptsize$\pm$.03} & 0.91{\scriptsize$\pm$.03} & 0.88{\scriptsize$\pm$.01} & 0.65{\scriptsize$\pm$.02} & 0.93{\scriptsize$\pm$.01} & 0.61{\scriptsize$\pm$.01} & 0.33{\scriptsize$\pm$.07} \\
\midrule
VISam G& 0.87{\scriptsize$\pm$.27} & 0.79{\scriptsize$\pm$.03} & 0.87{\scriptsize$\pm$.01} & 0.85{\scriptsize$\pm$.01} & 0.69{\scriptsize$\pm$.02} & 0.96{\scriptsize$\pm$.01} & 0.72{\scriptsize$\pm$.03} & 0.70{\scriptsize$\pm$.08} \\
 Cluster & 1.00 & 0.83{\scriptsize$\pm$.02} & 0.90{\scriptsize$\pm$.02} & 0.88{\scriptsize$\pm$.01} & 0.73{\scriptsize$\pm$.02} & 0.96{\scriptsize$\pm$.01} & 0.72{\scriptsize$\pm$.01} & 0.79{\scriptsize$\pm$.19} \\
\midrule
VISpm G& 0.20{\scriptsize$\pm$.24} & 0.81{\scriptsize$\pm$.03} & 0.90{\scriptsize$\pm$.01} & 0.89{\scriptsize$\pm$.01} & 0.74{\scriptsize$\pm$.03} & 0.94{\scriptsize$\pm$.01} & 0.71{\scriptsize$\pm$.03} & 0.49{\scriptsize$\pm$.16} \\
 Cluster & 0.00 & 0.87{\scriptsize$\pm$.01} & 0.91{\scriptsize$\pm$.01} & 0.90{\scriptsize$\pm$.01} & 0.79{\scriptsize$\pm$.01} & 0.94{\scriptsize$\pm$.01} & 0.70{\scriptsize$\pm$.03} & 0.41{\scriptsize$\pm$.10} \\
\midrule
VISl G& 0.40{\scriptsize$\pm$.49} & 0.73{\scriptsize$\pm$.06} & 0.89{\scriptsize$\pm$.03} & 0.89{\scriptsize$\pm$.02} & 0.66{\scriptsize$\pm$.05} & 0.96{\scriptsize$\pm$.01} & 0.71{\scriptsize$\pm$.04} & 0.71{\scriptsize$\pm$.18} \\
 Cluster & 0.00 & 0.79{\scriptsize$\pm$.03} & 0.90{\scriptsize$\pm$.01} & 0.88{\scriptsize$\pm$.01} & 0.72{\scriptsize$\pm$.02} & 0.96{\scriptsize$\pm$.01} & 0.73{\scriptsize$\pm$.02} & 0.63{\scriptsize$\pm$.18} \\
\midrule
V1 NV& -- & 0.99{\scriptsize$\pm$.01} & 0.99{\scriptsize$\pm$.01} & 0.98{\scriptsize$\pm$.01} & 0.94{\scriptsize$\pm$.01} & 0.86{\scriptsize$\pm$.01} & 0.90{\scriptsize$\pm$.01} & 1.00 \\
 Cluster & -- & 0.98{\scriptsize$\pm$.01} & 0.98{\scriptsize$\pm$.01} & 0.97{\scriptsize$\pm$.01} & 0.94{\scriptsize$\pm$.01} & 0.87{\scriptsize$\pm$.01} & 0.90{\scriptsize$\pm$.01} & 1.00 \\
\midrule
VISam NV& -- & 0.99{\scriptsize$\pm$.00} & 0.99{\scriptsize$\pm$.00} & 0.99{\scriptsize$\pm$.00} & 0.96{\scriptsize$\pm$.00} & 0.84{\scriptsize$\pm$.01} & 0.90{\scriptsize$\pm$.01} & 0.96{\scriptsize$\pm$.04} \\
 Cluster & -- & 0.99{\scriptsize$\pm$.00} & 1.00{\scriptsize$\pm$.00} & 0.99{\scriptsize$\pm$.00} & 0.97{\scriptsize$\pm$.00} & 0.86{\scriptsize$\pm$.01} & 0.90{\scriptsize$\pm$.01} & 1.00 \\
\midrule
VISrl NV& -- & 0.99{\scriptsize$\pm$.01} & 0.99{\scriptsize$\pm$.01} & 0.99{\scriptsize$\pm$.00} & 0.96{\scriptsize$\pm$.00} & 0.84{\scriptsize$\pm$.01} & 0.88{\scriptsize$\pm$.01} & 1.00 \\
 Cluster & -- & 0.99{\scriptsize$\pm$.00} & 0.99{\scriptsize$\pm$.00} & 0.99{\scriptsize$\pm$.00} & 0.96{\scriptsize$\pm$.00} & 0.84{\scriptsize$\pm$.02} & 0.88{\scriptsize$\pm$.01} & 1.00 \\
\midrule
VISpm NV& -- & 0.99{\scriptsize$\pm$.00} & 0.99{\scriptsize$\pm$.00} & 0.99{\scriptsize$\pm$.00} & 0.96{\scriptsize$\pm$.01} & 0.85{\scriptsize$\pm$.01} & 0.90{\scriptsize$\pm$.01} & 0.96{\scriptsize$\pm$.05} \\
 Cluster & -- & 0.99{\scriptsize$\pm$.01} & 0.99{\scriptsize$\pm$.01} & 0.99{\scriptsize$\pm$.01} & 0.96{\scriptsize$\pm$.01} & 0.84{\scriptsize$\pm$.03} & 0.89{\scriptsize$\pm$.02} & 0.94{\scriptsize$\pm$.08} \\
\midrule
VISl NV& -- & 0.98{\scriptsize$\pm$.01} & 0.97{\scriptsize$\pm$.02} & 0.96{\scriptsize$\pm$.01} & 0.94{\scriptsize$\pm$.01} & 0.84{\scriptsize$\pm$.01} & 0.89{\scriptsize$\pm$.01} & 0.74{\scriptsize$\pm$.04} \\
 Cluster & -- & 0.97{\scriptsize$\pm$.03} & 0.96{\scriptsize$\pm$.04} & 0.95{\scriptsize$\pm$.02} & 0.94{\scriptsize$\pm$.01} & 0.84{\scriptsize$\pm$.01} & 0.88{\scriptsize$\pm$.01} & 0.70{\scriptsize$\pm$.05} \\
\midrule
V1 F& 0.87{\scriptsize$\pm$.05} & 0.91{\scriptsize$\pm$.02} & 0.92{\scriptsize$\pm$.02} & 0.86{\scriptsize$\pm$.02} & 0.84{\scriptsize$\pm$.03} & 0.90{\scriptsize$\pm$.02} & 0.87{\scriptsize$\pm$.01} & 0.88{\scriptsize$\pm$.14} \\
 Cluster & 0.88{\scriptsize$\pm$.03} & 0.93{\scriptsize$\pm$.01} & 0.93{\scriptsize$\pm$.01} & 0.88{\scriptsize$\pm$.01} & 0.86{\scriptsize$\pm$.01} & 0.93{\scriptsize$\pm$.01} & 0.88{\scriptsize$\pm$.00} & 0.87{\scriptsize$\pm$.17} \\
\midrule
Retina& 0.98{\scriptsize$\pm$.02} & 0.89{\scriptsize$\pm$.03} & 0.97{\scriptsize$\pm$.01} & 0.96{\scriptsize$\pm$.01} & 0.87{\scriptsize$\pm$.01} & 0.94{\scriptsize$\pm$.00} & 0.93{\scriptsize$\pm$.01} & 0.81{\scriptsize$\pm$.11} \\
 Cluster & 0.99{\scriptsize$\pm$.01} & 0.95{\scriptsize$\pm$.03} & 0.98{\scriptsize$\pm$.02} & 0.97{\scriptsize$\pm$.01} & 0.91{\scriptsize$\pm$.01} & 0.93{\scriptsize$\pm$.01} & 0.92{\scriptsize$\pm$.00} & 0.82{\scriptsize$\pm$.10} \\
\midrule
FNN& 0.92{\scriptsize$\pm$.06} & 0.88{\scriptsize$\pm$.04} & 0.94{\scriptsize$\pm$.02} & 0.89{\scriptsize$\pm$.01} & 0.80{\scriptsize$\pm$.03} & 0.97{\scriptsize$\pm$.00} & 0.87{\scriptsize$\pm$.01} & 0.63{\scriptsize$\pm$.24} \\
 Cluster & 0.95{\scriptsize$\pm$.06} & 0.75{\scriptsize$\pm$.09} & 0.85{\scriptsize$\pm$.08} & 0.83{\scriptsize$\pm$.04} & 0.71{\scriptsize$\pm$.06} & 0.97{\scriptsize$\pm$.00} & 0.89{\scriptsize$\pm$.01} & 0.60{\scriptsize$\pm$.29} \\
\midrule
Flyvision& 0.95{\scriptsize$\pm$.04} & 0.96{\scriptsize$\pm$.00} & 0.94{\scriptsize$\pm$.00} & 0.85{\scriptsize$\pm$.01} & 0.94{\scriptsize$\pm$.00} & 0.82{\scriptsize$\pm$.01} & 0.89{\scriptsize$\pm$.00} & 1.00 \\
 Cluster & 0.98{\scriptsize$\pm$.01} & 0.94{\scriptsize$\pm$.00} & 0.88{\scriptsize$\pm$.02} & 0.82{\scriptsize$\pm$.01} & 0.92{\scriptsize$\pm$.01} & 0.80{\scriptsize$\pm$.01} & 0.87{\scriptsize$\pm$.01} & 0.98{\scriptsize$\pm$.03} \\
\midrule
R2+1D L4& 1.00{\scriptsize$\pm$.01} & 0.82{\scriptsize$\pm$.04} & 0.93{\scriptsize$\pm$.01} & 0.93{\scriptsize$\pm$.01} & 0.84{\scriptsize$\pm$.03} & 0.73{\scriptsize$\pm$.04} & 0.92{\scriptsize$\pm$.01} & 0.69{\scriptsize$\pm$.06} \\
 Cluster & 1.00 & 0.76{\scriptsize$\pm$.06} & 0.89{\scriptsize$\pm$.02} & 0.88{\scriptsize$\pm$.01} & 0.78{\scriptsize$\pm$.04} & 0.69{\scriptsize$\pm$.02} & 0.90{\scriptsize$\pm$.01} & 0.68{\scriptsize$\pm$.06} \\
\midrule
ViT & 0.90{\scriptsize$\pm$.07} & 0.94{\scriptsize$\pm$.02} & 0.85{\scriptsize$\pm$.08} & 0.82{\scriptsize$\pm$.04} & 0.73{\scriptsize$\pm$.08} & 0.47{\scriptsize$\pm$.07} & 0.76{\scriptsize$\pm$.01} & 1.00 \\
 Clustered & 0.87{\scriptsize$\pm$.04} & 0.98{\scriptsize$\pm$.00} & 0.96{\scriptsize$\pm$.01} & 0.87{\scriptsize$\pm$.01} & 0.87{\scriptsize$\pm$.02} & 0.60{\scriptsize$\pm$.05} & 0.80{\scriptsize$\pm$.01} & 1.00 \\
\midrule
Raptor & 0.84{\scriptsize$\pm$.02} & 0.65{\scriptsize$\pm$.01} & 0.83{\scriptsize$\pm$.00} & 0.70{\scriptsize$\pm$.00} & 0.56{\scriptsize$\pm$.01} & 1.00{\scriptsize$\pm$.00} & 0.99{\scriptsize$\pm$.00} & 0.37{\scriptsize$\pm$.03} \\
 Clustered & 0.86{\scriptsize$\pm$.01} & 0.69{\scriptsize$\pm$.00} & 0.84{\scriptsize$\pm$.00} & 0.70{\scriptsize$\pm$.00} & 0.57{\scriptsize$\pm$.01} & 1.00{\scriptsize$\pm$.00} & 0.98{\scriptsize$\pm$.00} & 0.30{\scriptsize$\pm$.03} \\
\bottomrule
\end{tabular}
\end{table*}

\subsection{FPS Subpopulation Showcases}
\label{app:fps_showcases}
\begin{figure*}[p]
    \centering
    \includegraphics[width=0.9\linewidth]{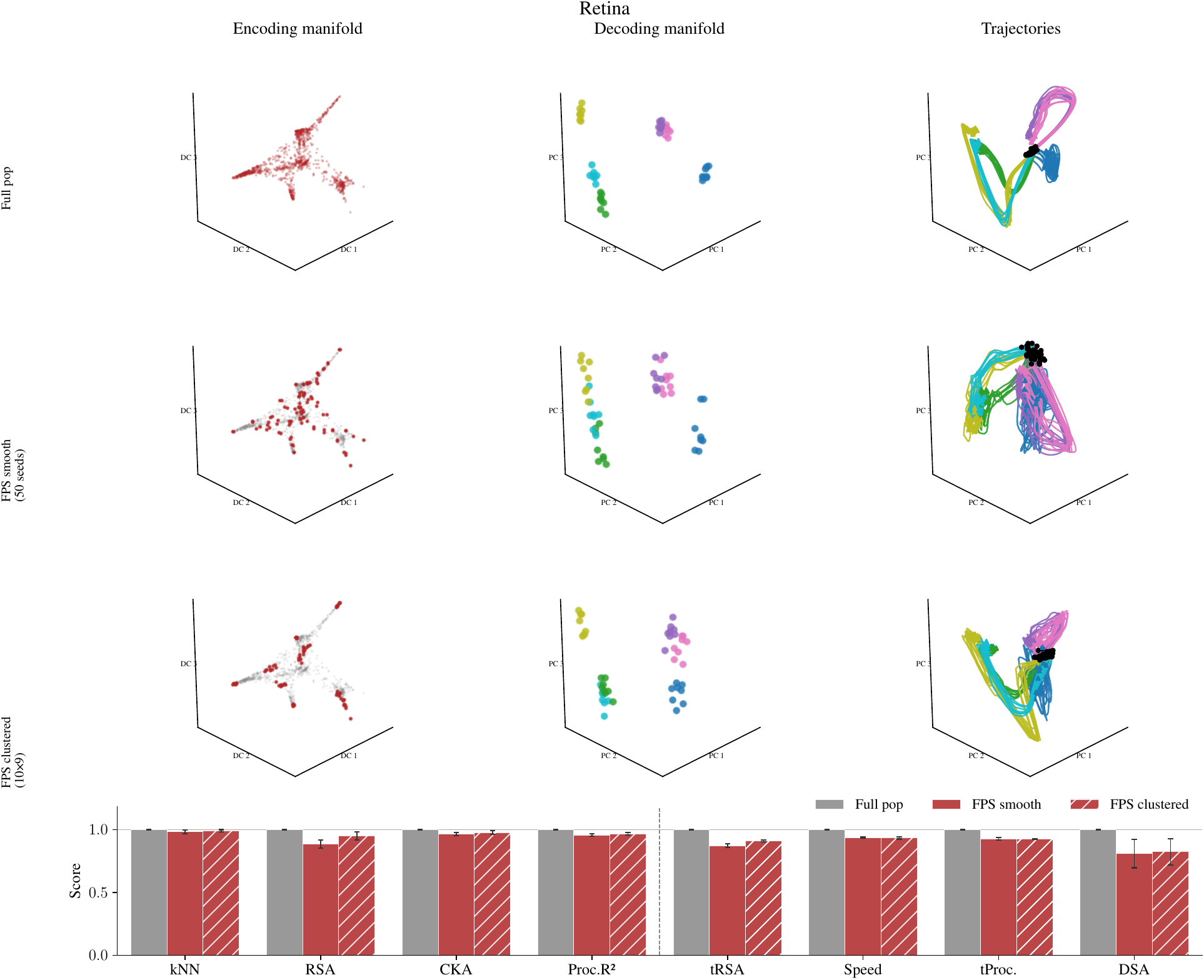}
    \caption{\textbf{Retina --- FPS subpopulation showcase.} Each row shows the 3-D encoding manifold (left), 3-D decoding manifold (center), and stimulus trajectories (right) for the full population (top), FPS continuous subpopulation ($n{=}50$ seeds, $m{=}1$ neighbor, $\approx$100 neurons; middle), and FPS clustered subpopulation ($n{=}10$ seeds, $m{=}9$ neighbors, $\approx$100 neurons; bottom). The bottom bar chart reports all eight decoding metrics relative to the full population (mean\,$\pm$\,std over five random seeds $\{42\text{--}46\}$).}
    \label{fig:retina_fps}
\end{figure*}

\begin{figure*}[p]
    \centering
    \includegraphics[width=0.9\linewidth]{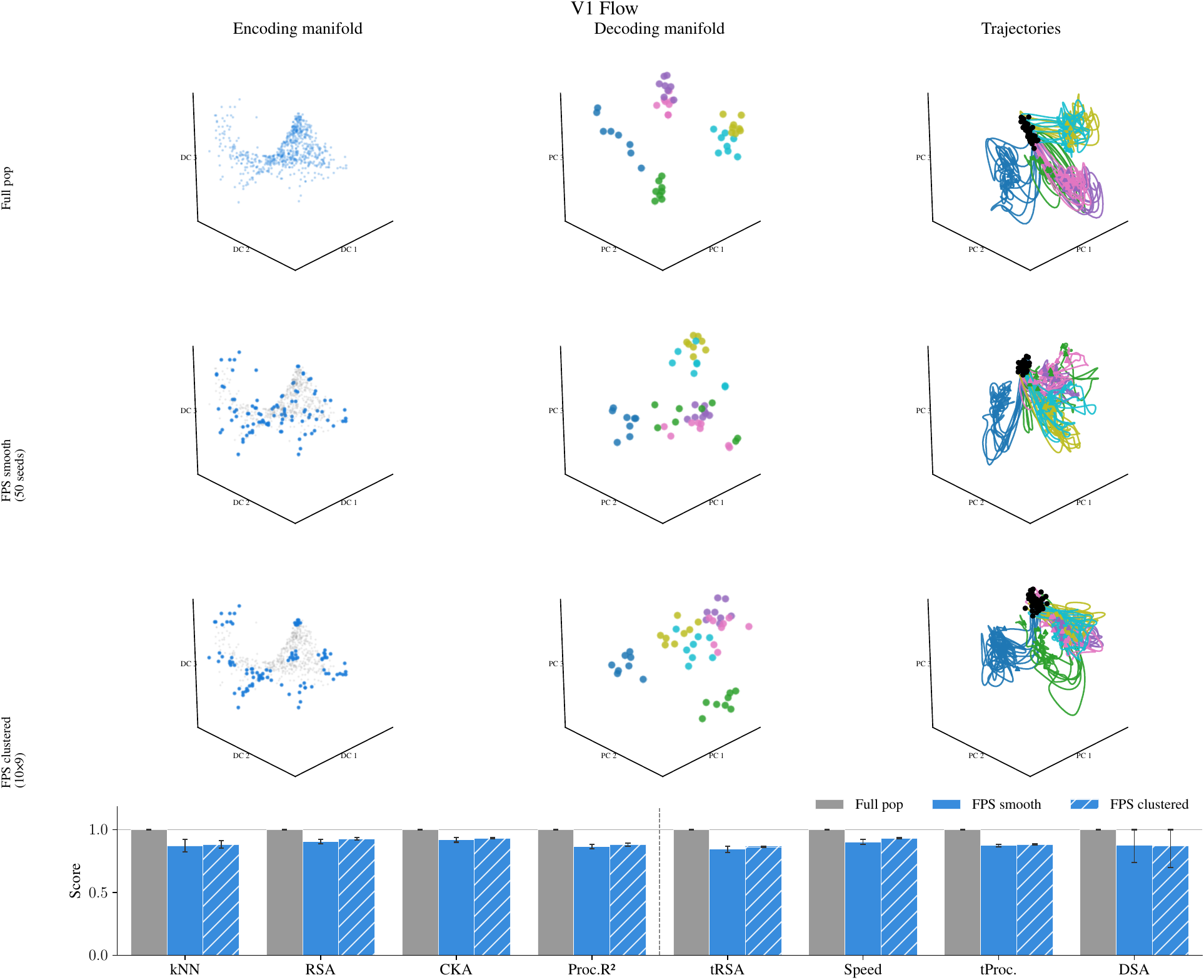}
    \caption{\textbf{V1 --- FPS subpopulation showcase.} Layout as in Figure~\ref{fig:retina_fps}. FPS continuous: $n{=}50$, $m{=}1$; FPS clustered: $n{=}10$, $m{=}9$. Both yield $\approx$100 neurons.}
    \label{fig:v1_fps}
\end{figure*}

\begin{figure*}[p]
    \centering
    \includegraphics[width=0.9\linewidth]{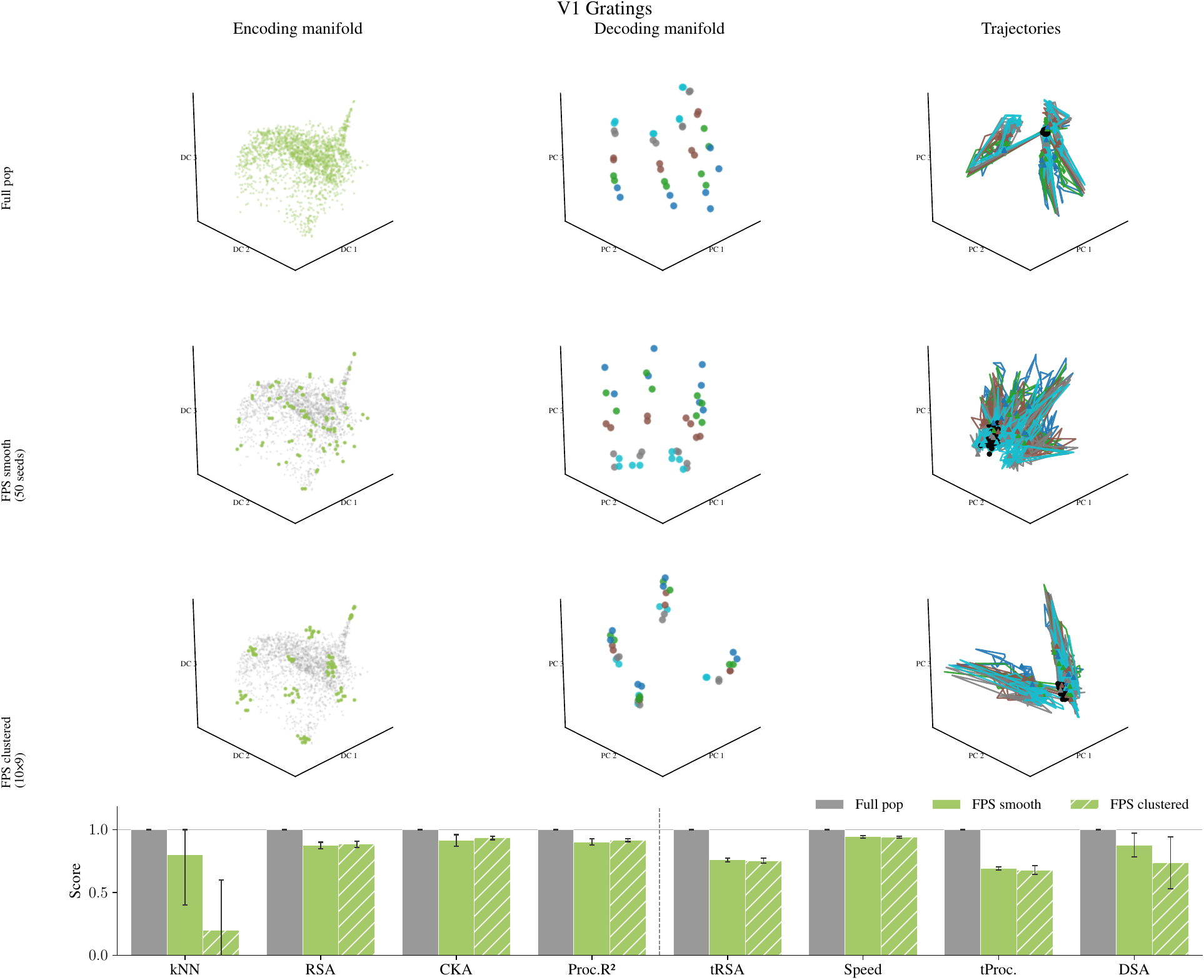}
    \caption{\textbf{Allen VISp (drifting gratings) --- FPS subpopulation showcase.} Layout as in Figure~\ref{fig:retina_fps}. FPS continuous: $n{=}50$, $m{=}1$; FPS clustered: $n{=}10$, $m{=}9$.}
    \label{fig:allen_visp_fps}
\end{figure*}

\begin{figure*}[p]
    \centering
    \includegraphics[width=0.9\linewidth]{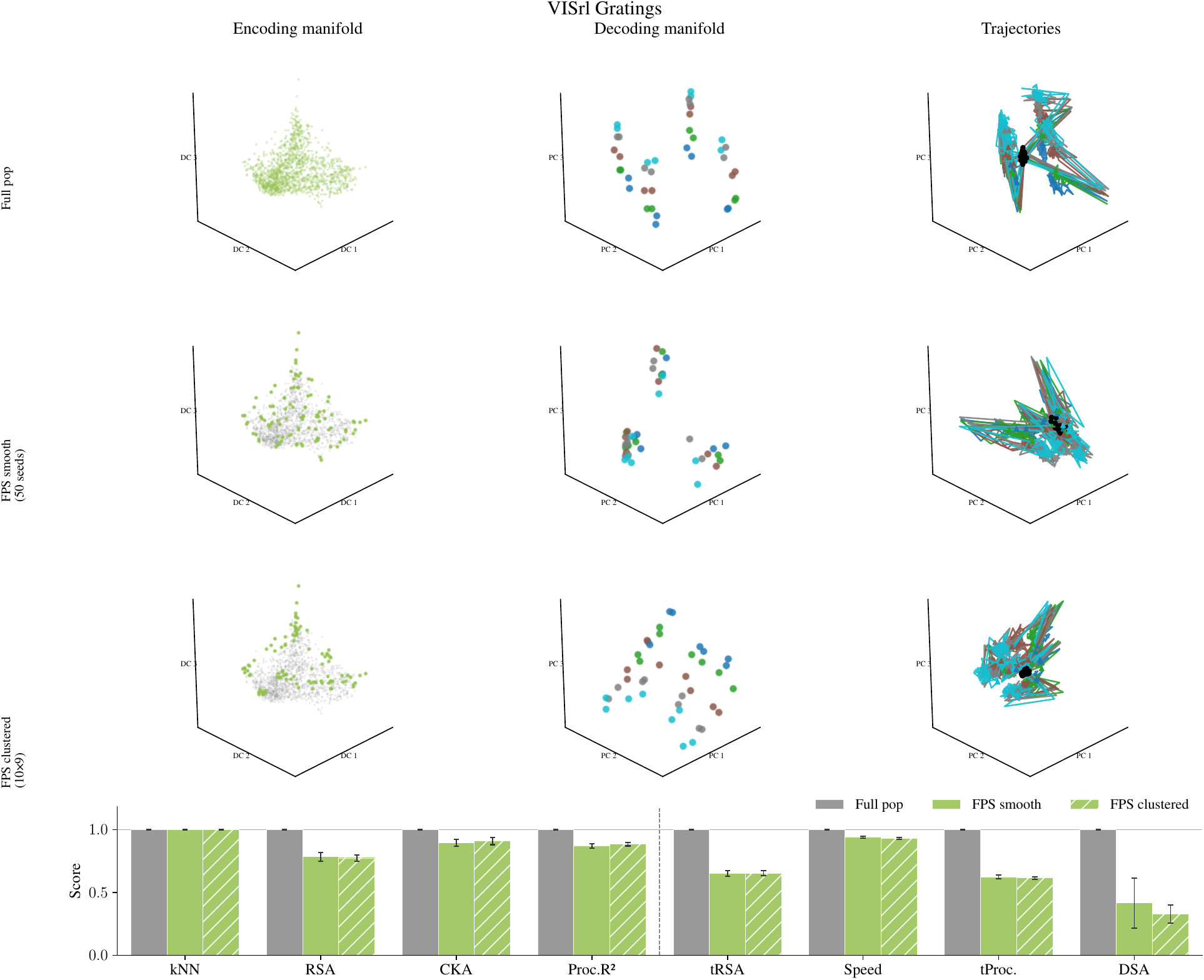}
    \caption{\textbf{Allen VISrl (drifting gratings) --- FPS subpopulation showcase.} Layout as in Figure~\ref{fig:retina_fps}.}
    \label{fig:allen_visrl_fps}
\end{figure*}

\begin{figure*}[p]
    \centering
    \includegraphics[width=0.9\linewidth]{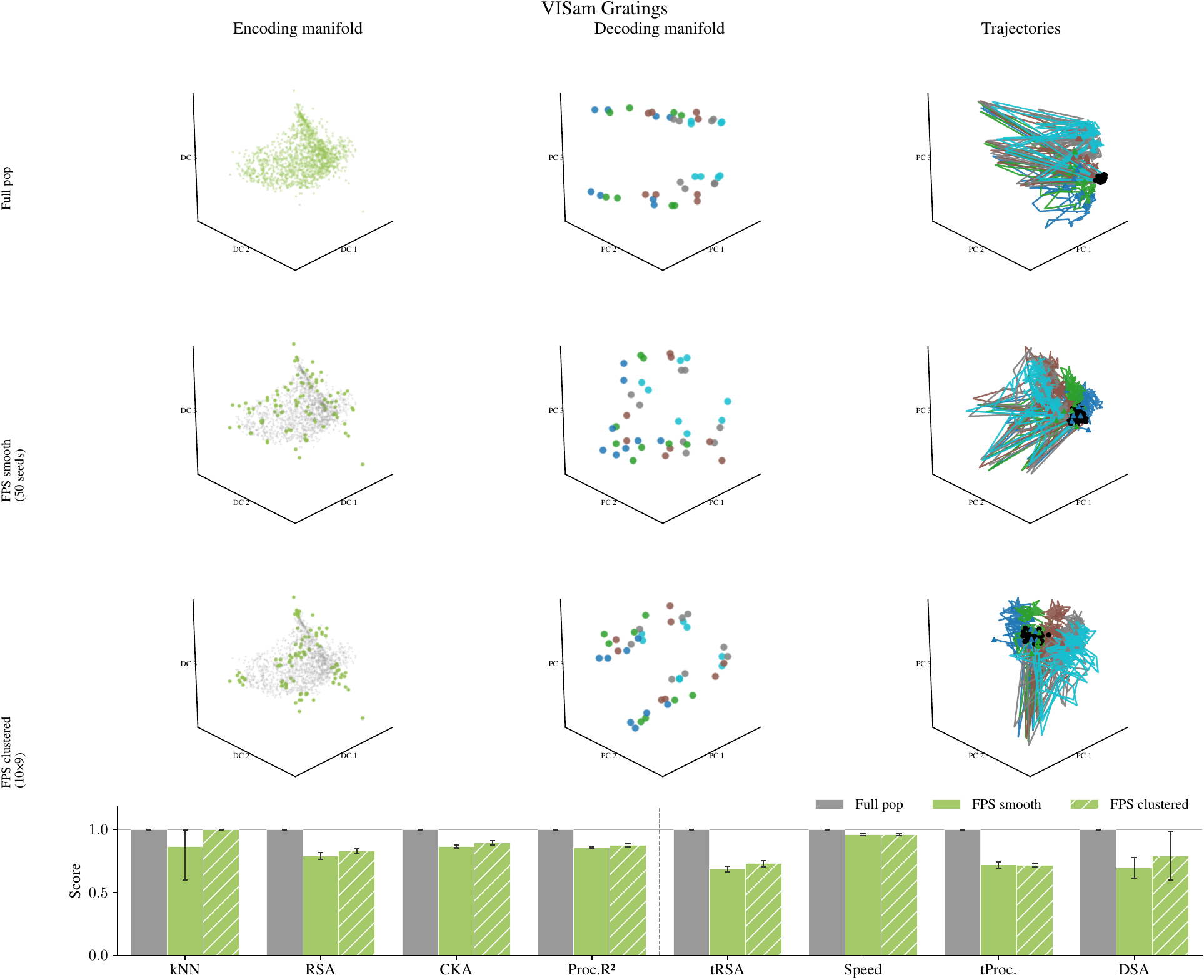}
    \caption{\textbf{Allen VISam (drifting gratings) --- FPS subpopulation showcase.} Layout as in Figure~\ref{fig:retina_fps}.}
    \label{fig:allen_visam_fps}
\end{figure*}

\begin{figure*}[p]
    \centering
    \includegraphics[width=0.9\linewidth]{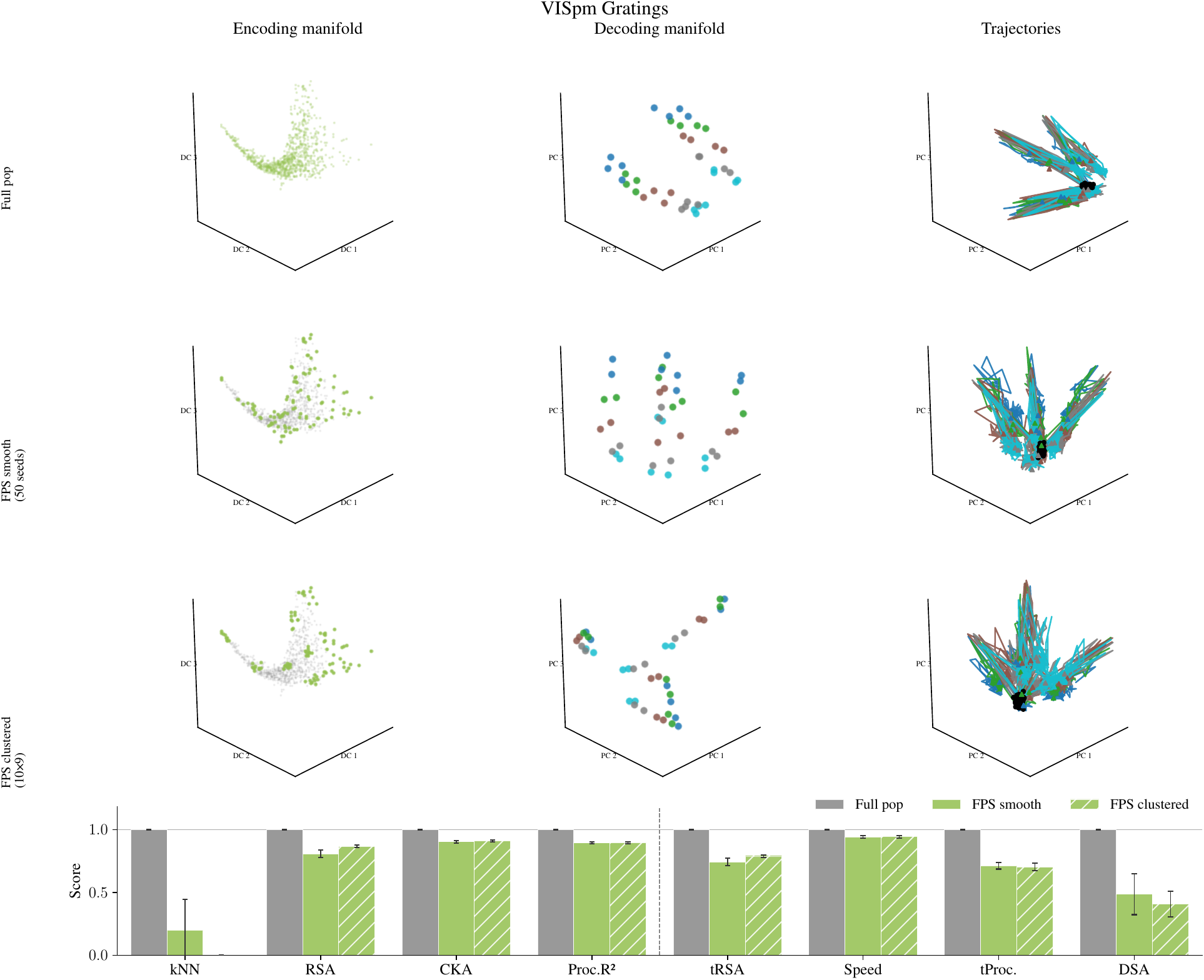}
    \caption{\textbf{Allen VISpm (drifting gratings) --- FPS subpopulation showcase.} Layout as in Figure~\ref{fig:retina_fps}.}
    \label{fig:allen_vispm_fps}
\end{figure*}

\begin{figure*}[p]
    \centering
    \includegraphics[width=0.9\linewidth]{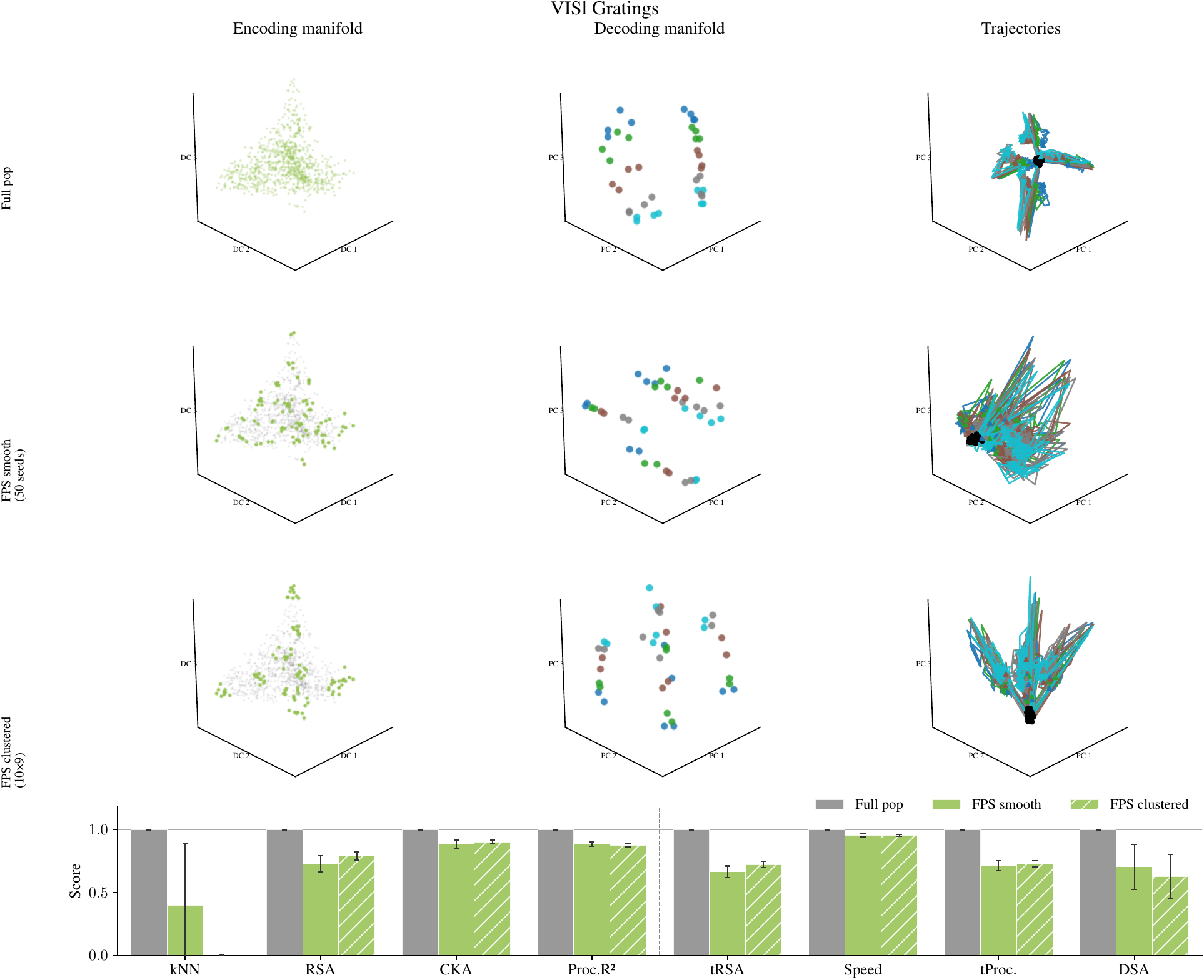}
    \caption{\textbf{Allen VISl (drifting gratings) --- FPS subpopulation showcase.} Layout as in Figure~\ref{fig:retina_fps}.}
    \label{fig:allen_visl_fps}
\end{figure*}

\begin{figure*}[p]
    \centering
    \includegraphics[width=0.9\linewidth]{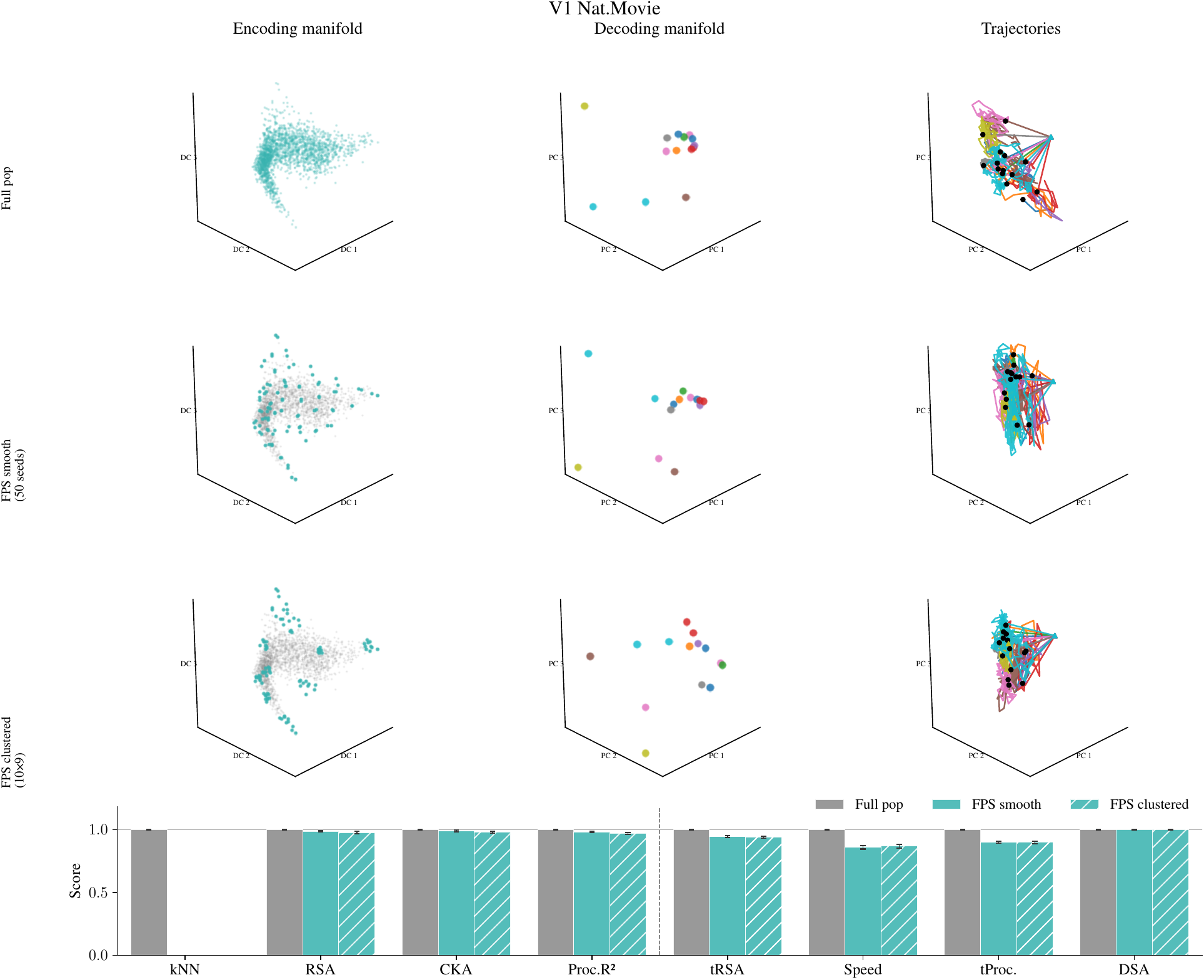}
    \caption{\textbf{Allen VISp (natural movie, 156 scenes) --- FPS subpopulation showcase.} Layout as in Figure~\ref{fig:retina_fps}. Because the natural movie stimulus has a single direction ($D{=}1$), k-NN accuracy is not computed (marked ``--'' in the bar chart).}
    \label{fig:nat_visp_fps}
\end{figure*}

\begin{figure*}[p]
    \centering
    \includegraphics[width=0.9\linewidth]{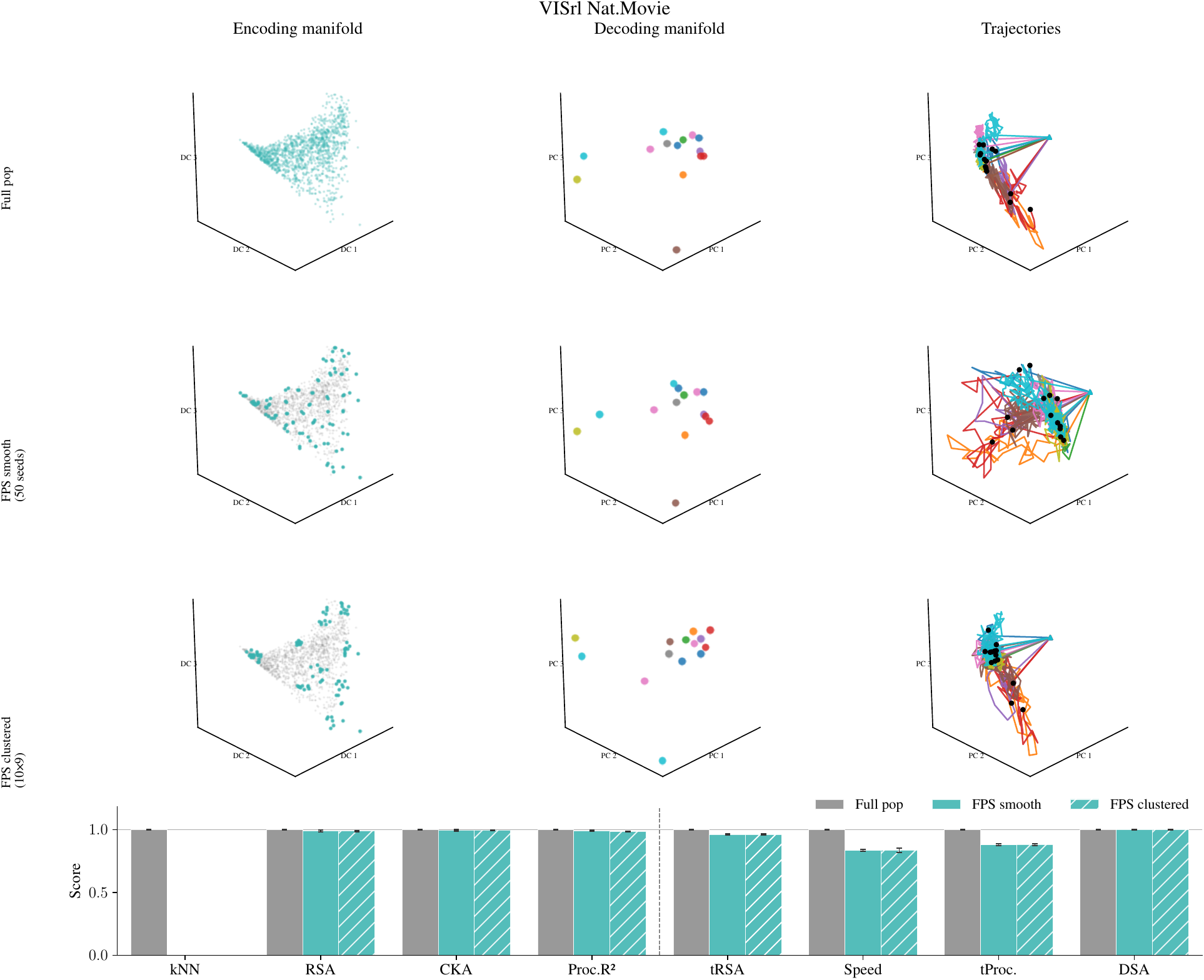}
    \caption{\textbf{Allen VISrl (natural movie) --- FPS subpopulation showcase.} Layout as in Figure~\ref{fig:nat_visp_fps}.}
    \label{fig:nat_visrl_fps}
\end{figure*}

\begin{figure*}[p]
    \centering
    \includegraphics[width=0.9\linewidth]{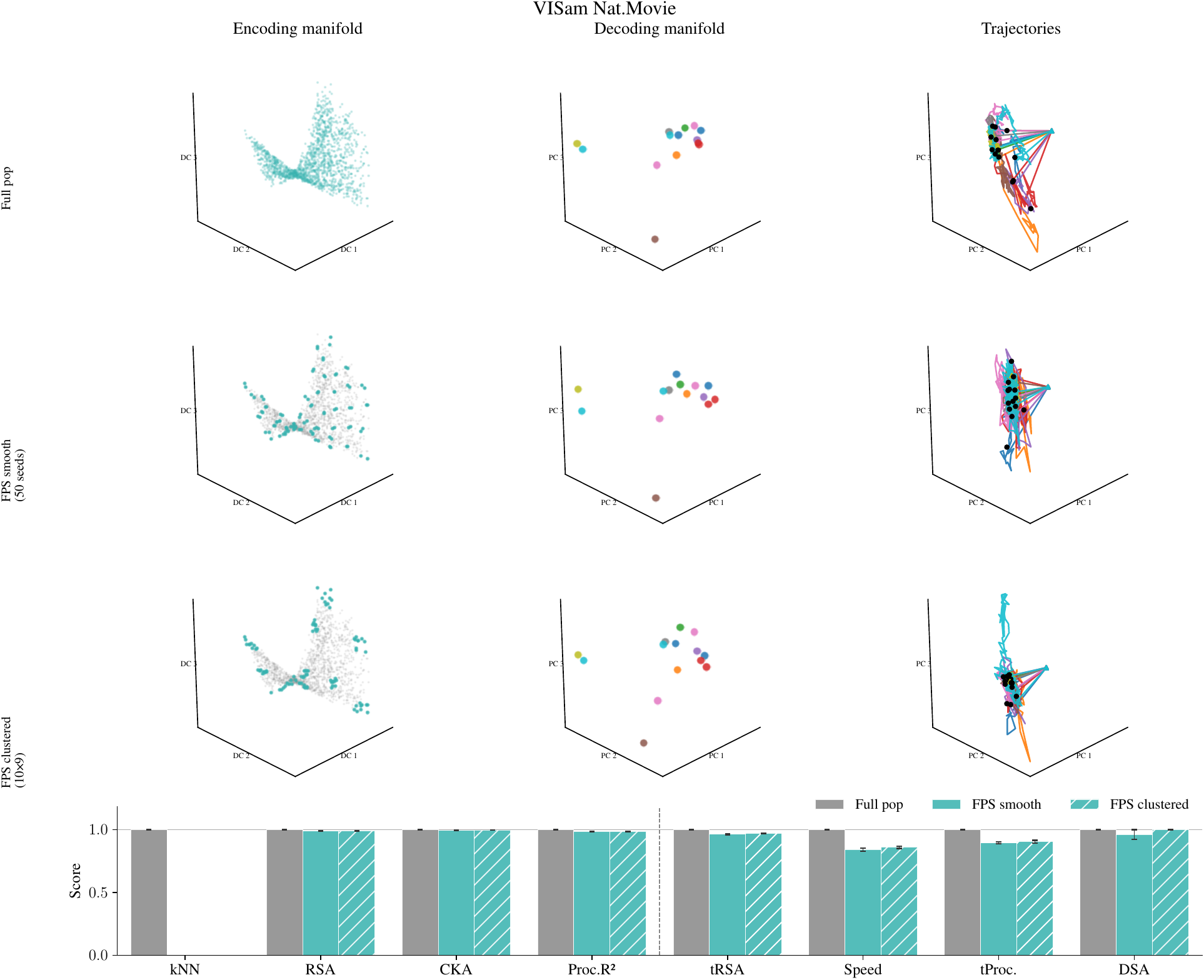}
    \caption{\textbf{Allen VISam (natural movie) --- FPS subpopulation showcase.} Layout as in Figure~\ref{fig:nat_visp_fps}.}
    \label{fig:nat_visam_fps}
\end{figure*}

\begin{figure*}[p]
    \centering
    \includegraphics[width=0.9\linewidth]{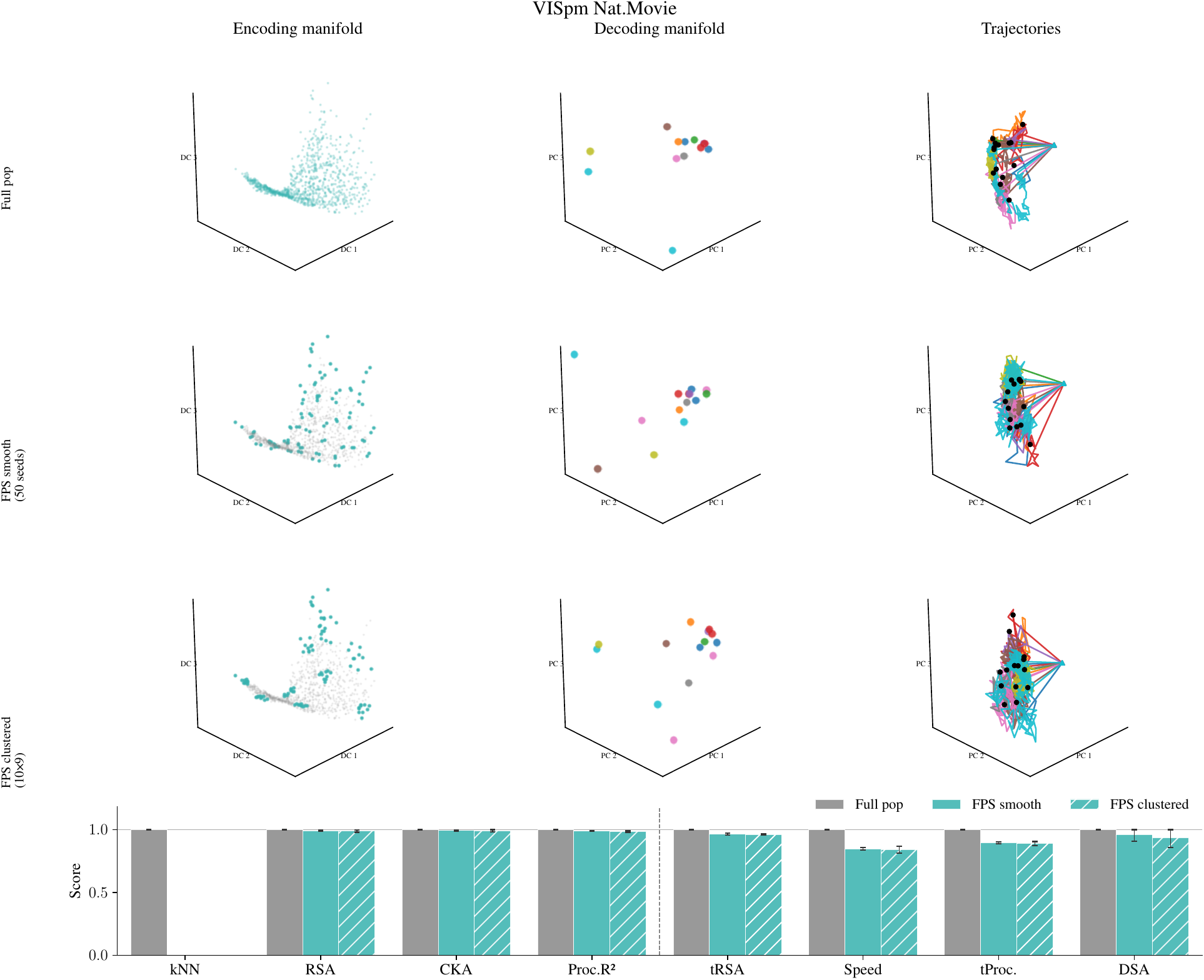}
    \caption{\textbf{Allen VISpm (natural movie) --- FPS subpopulation showcase.} Layout as in Figure~\ref{fig:nat_visp_fps}.}
    \label{fig:nat_vispm_fps}
\end{figure*}

\begin{figure*}[p]
    \centering
    \includegraphics[width=0.9\linewidth]{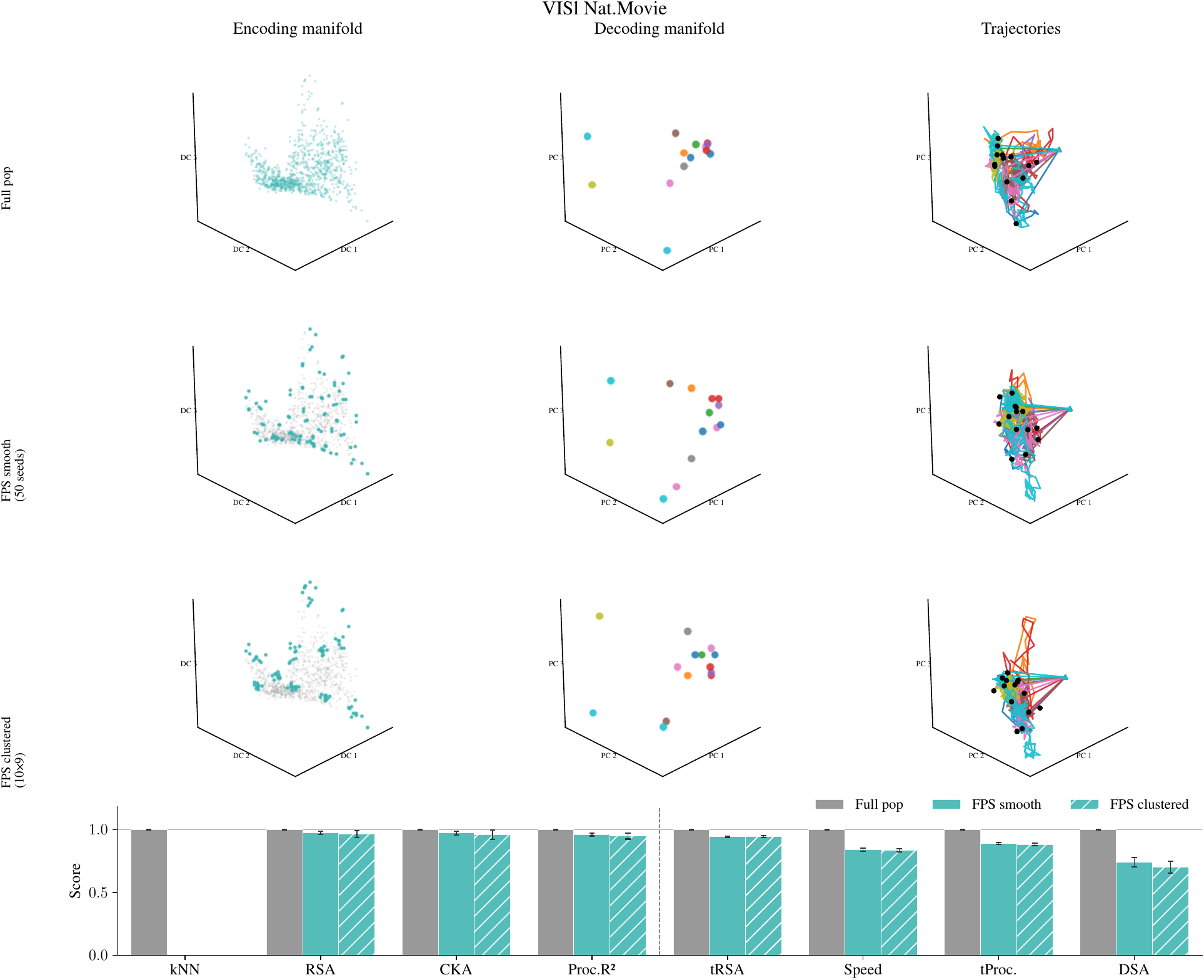}
    \caption{\textbf{Allen VISl (natural movie) --- FPS subpopulation showcase.} Layout as in Figure~\ref{fig:nat_visp_fps}.}
    \label{fig:nat_visl_fps}
\end{figure*}

\begin{figure*}[p]
    \centering
    \includegraphics[width=0.9\linewidth]{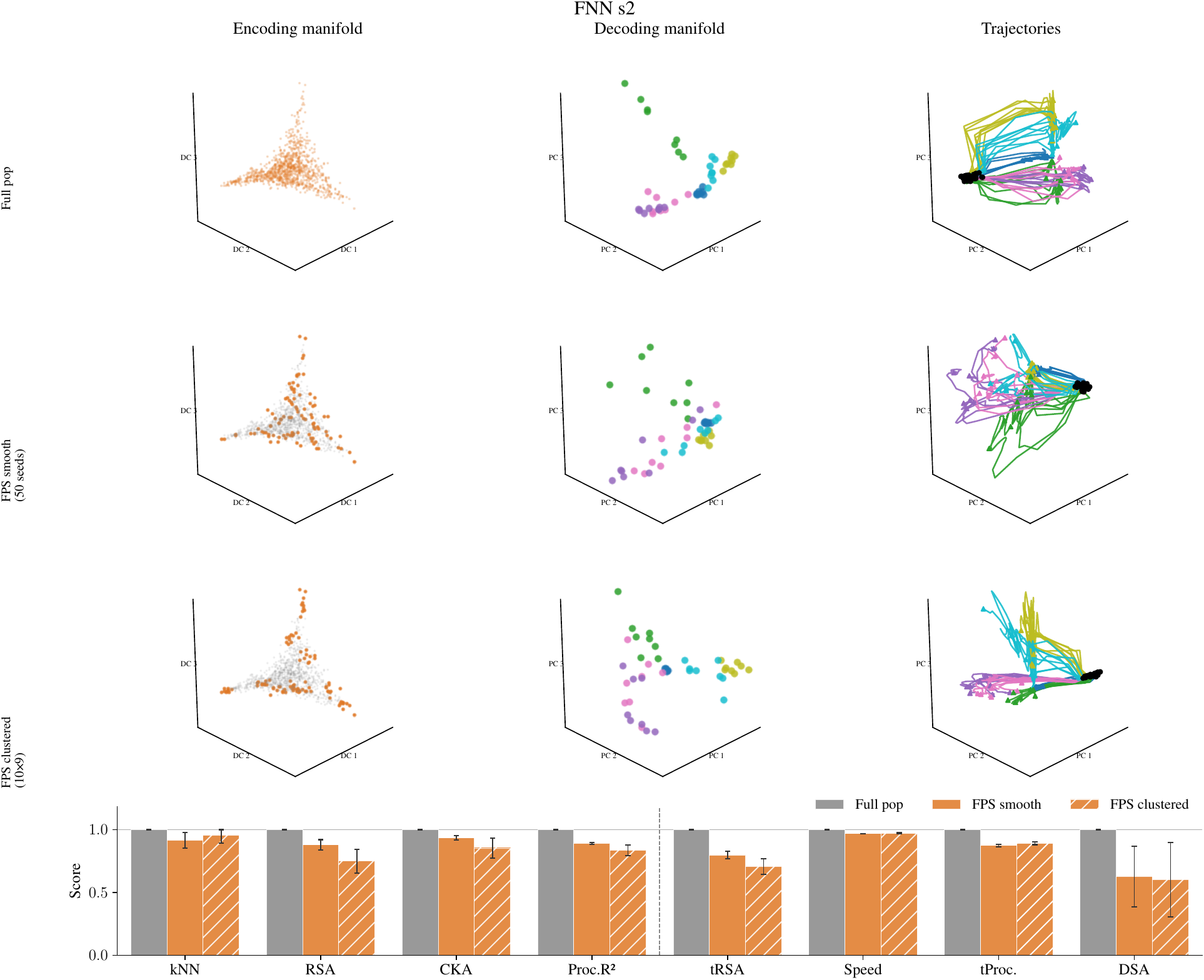}
    \caption{\textbf{FNN output layer --- FPS subpopulation showcase.} Layout as in Figure~\ref{fig:retina_fps}. 2000 sampled neurons from the FNN recurrent output layer, probed with the 6-class $\times$ 8-direction optical-flow stimulus ensemble.}
    \label{fig:fnn_out_fps}
\end{figure*}

\begin{figure*}[p]
    \centering
    \includegraphics[width=0.9\linewidth]{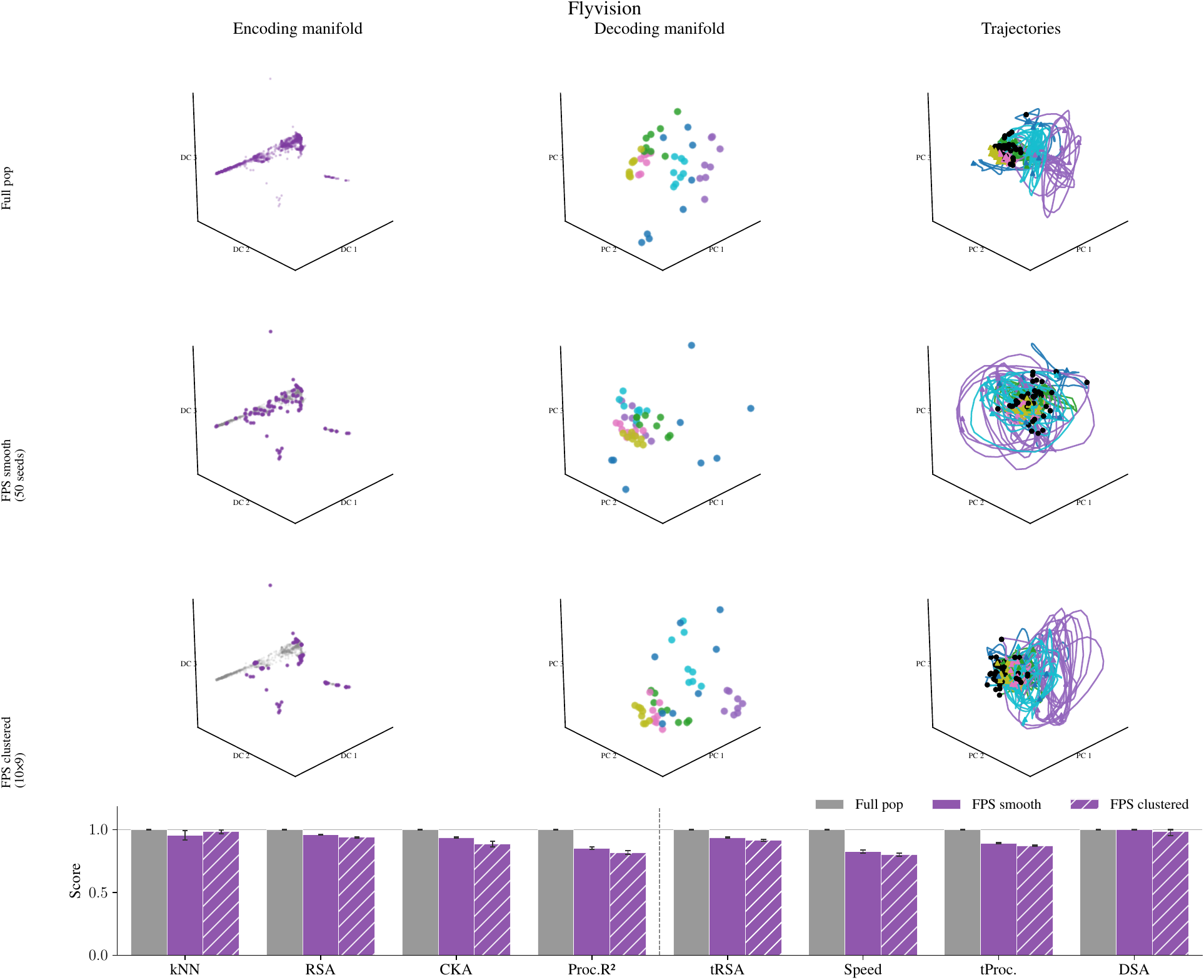}
    \caption{\textbf{Flyvision (T, Tm cells) --- FPS subpopulation showcase.} Layout as in Figure~\ref{fig:retina_fps}. 1700 T and Tm cell types from a connectome-constrained fly visual system model \citep{lappalainenConnectomeconstrainedNetworksPredict2024}.}
    \label{fig:flyvis_fps}
\end{figure*}

\begin{figure*}[p]
    \centering
    \includegraphics[width=0.9\linewidth]{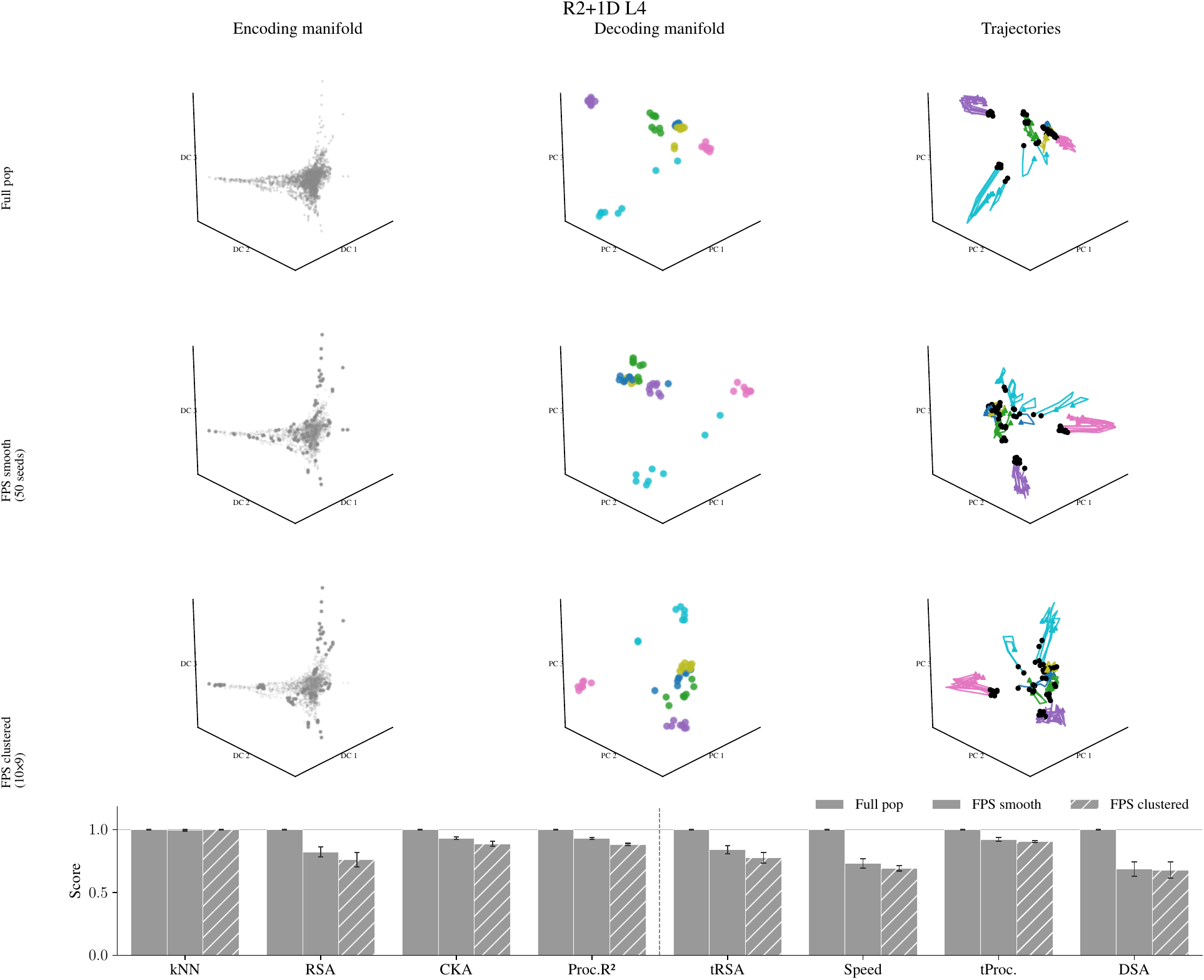}
    \caption{\textbf{R(2+1)D layer 4 --- FPS subpopulation showcase.} Layout as in Figure~\ref{fig:retina_fps}. 1960 activations from layer 4 of a pretrained spatiotemporal video CNN \citep{tranCloserLookSpatiotemporal2018}.}
    \label{fig:r2plus1d_fps}
\end{figure*}

Figures~\ref{fig:retina_fps}--\ref{fig:r2plus1d_fps} show encoding manifold, decoding manifold, and stimulus trajectory visualizations for FPS continuous and clustered subpopulations across all biological systems and machine learning models, together with the full per-dataset decoding metric bar charts. Table~\ref{tab:showcase} reports all eight metrics numerically across all 17 datasets.

\subsubsection{Biological Systems}
\label{app:bio_fps}

\subsubsection{Machine Learning Models}
\label{app:ml_fps}

\begin{figure*}[p]
    \centering
    \includegraphics[width=0.9\linewidth]{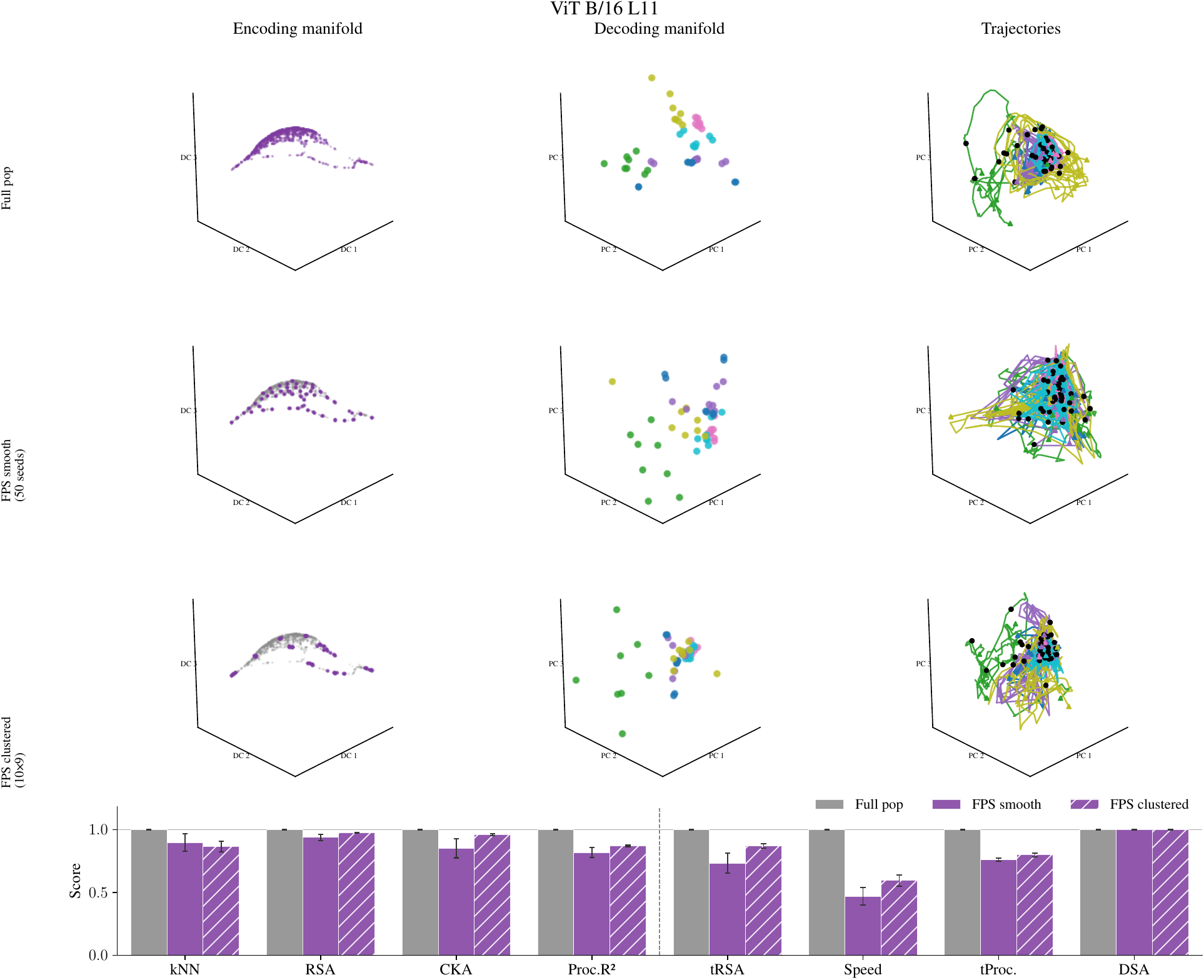}
    \caption{\textbf{ViT Layer 11 --- FPS subpopulation showcase.} Layout as in Figure~\ref{fig:retina_fps}. Activations from a pretrained vision transformer \citep{dosovitskiyImageWorth16x162021}.}
    \label{fig:vit}
\end{figure*}

\begin{figure*}[p]
    \centering
    \includegraphics[width=0.9\linewidth]{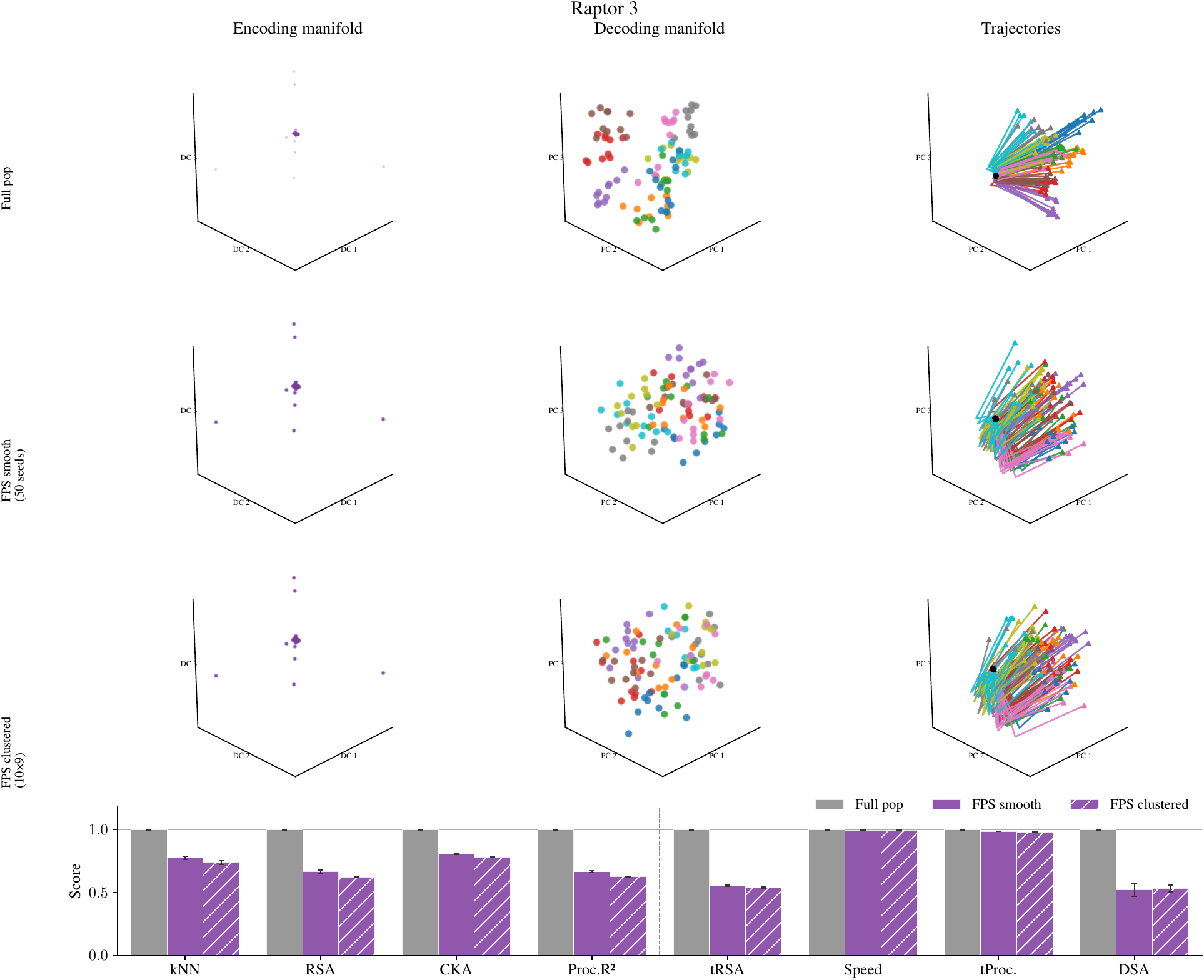}
    \caption{\textbf{Raptor --- FPS subpopulation showcase.} Layout as in Figure~\ref{fig:retina_fps}. Activations from a pretrained recurrent vision transformer \citep{jacobsBlockRecurrentDynamicsVision2026}.}
    \label{fig:raptor}
\end{figure*}

\section{Methods}
\label{app:methods}

This section provides full methodological details for all experiments, datasets, and pipelines used in this paper.

\subsection{Data}
\label{app:data}

\subsubsection{Biological data: Retina and V1}
Retinal and V1 neural recordings are taken from \citet{dyballaPopulationEncodingStimulus2024}. The stimulus ensemble consists of 88 unique sequences of drifting square-wave gratings and optical flows moving in 8 directions, grouped into 6 base-stimulus classes at two spatial frequencies (medium and high). For each neuron, the spatial frequency eliciting the larger population response is selected. The response tensor has shape $(N \times S \times K \times T)$, where $N$ is the number of neurons (Retina: $N=1{,}139$; V1: $N=630$, after outlier removal in the encoding pipeline), $S$ is the number of stimuli (6 after spatial-frequency selection), $K = 8$ is the number of movement directions, and $T$ is the number of time bins ($T = 135$ for biological data at $\sim 7.5$~ms resolution).

\subsubsection{Allen Brain Observatory}
We use Neuropixels electrophysiology data from the Allen Brain Observatory Visual Coding dataset \citep{devriesLargescaleStandardizedPhysiological2020}, accessed via the AllenSDK \texttt{EcephysProjectCache}. We select all 32 sessions of the \texttt{brain\_observatory\_1.1} type, which include full 8-direction $\times$ 5-temporal-frequency drifting grating sweeps ($\sim 15$ repeats per condition) and natural movie stimuli.

\paragraph{Drifting gratings.}
For each of the six cortical visual areas (VISp, VISl, VISam, VISpm, VISrl), units are pooled across sessions after quality filtering (ISI violations $< 0.5$, presence ratio $> 0.9$, amplitude $> 50~\mu$V). Trial-averaged PSTHs are computed in 50~ms bins over a 2-second window, yielding tensors of shape $(N_\text{area} \times 5 \times 8 \times 40)$, where the 5 stimuli correspond to temporal frequencies of 1, 2, 4, 8, and 15~Hz (spatial frequency fixed at 0.04~cpd, contrast 0.8).

\paragraph{Natural movie.}
Natural movie one (30~s clip, $\sim 900$ frames at 30~fps) is segmented into 156 non-overlapping scenes of $\sim 6$ frames each. Trial-averaged responses are computed per scene in the same 50~ms bins, yielding tensors of shape $(N_\text{area} \times 156 \times 1 \times T_\text{scene})$.

\subsubsection{Machine learning models}
We additionally analyze subpopulations in the following models:
\begin{itemize}
    \item \textbf{FNN} (Foundation Neural Network) \citep{wangFoundationModelNeural2025}: We sample 2000 neurons from the recurrent hidden state and from the output layer of the pretrained model, probed with the same 6-class $\times$ 8-direction stimulus ensemble as the biological data. Tensor shape: $(2000 \times 6 \times 8 \times 37)$.
    \item \textbf{Flyvision} \citep{lappalainenConnectomeconstrainedNetworksPredict2024}: T and Tm cell types from a connectome-constrained fly visual system model (1700 neurons, same stimulus protocol).
    \item \textbf{R(2+1)D} \citep{tranCloserLookSpatiotemporal2018}: Layer 4 activations of a pretrained spatiotemporal video network (1960 neurons).
    \item \textbf{ViT-B/16} \citep{dosovitskiyImageWorth16x162021}: A standard Vision Transformer with no recurrent structure. The same 37-frame pixel-roll sequences are used; frames are fed individually to the model and activations from a single encoder block are stacked to form the temporal axis. Each of the $197 = 1 + 196$ output tokens (CLS + patch tokens) at a given hidden dimension constitutes one ``neuron'' --- analogous to the feature-map $\times$ channel indexing used for convolutional models --- yielding up to $197 \times 768 = 151{,}296$ candidate units per layer, of which 2000 are retained by max-activation sampling (see Figure~\ref{fig:vit}).
    \item \textbf{Raptor} The pipeline also supports the alternative framing in which transformer \emph{layers} serve as the time axis \citep{jacobsBlockRecurrentDynamicsVision2026}; in that case single static frames (e.g.\ ImageNet images) are presented and activations are collected across all 12 encoder blocks, as done for the Raptor \citep{jacobsBlockRecurrentDynamicsVision2026} model (see Figure~\ref{fig:raptor}).
\end{itemize}

\subsection{Decoding manifold construction}
\label{app:decoding_pipeline}

The decoding manifold embeds stimulus conditions as points in neural activity space.

\paragraph{Preprocessing.}
Prior to all analyses, each response tensor is shifted to be non-negative: $X \leftarrow X - \min_\mathrm{nan}(X)$. Remaining NaN entries (variable-length padding for natural movie data) are replaced with 0 before PCA and metric computations.

\paragraph{Time-averaged manifold.}
Given the preprocessed population response tensor $X \in \mathbb{R}^{N \times S \times K \times T}$, we compute the time-averaged response matrix:
$$
\hat{X} \in \mathbb{R}^{SK \times N}, \qquad \hat{X}_{(s,d), n} = \frac{1}{T} \sum_{t=1}^{T} X_{n,s,d,t}.
$$
For datasets with NaN-padded variable-length stimuli, the mean is taken over valid (non-NaN) time bins only. PCA is applied to $\hat{X}$ to obtain a 3-component embedding $\mathbf{Z} = \text{PCA}_3(\hat{X}) \in \mathbb{R}^{SK \times 3}$ for visualization. k-NN accuracy, RSA, and CKA operate directly on the full-dimensional $\hat{X}$ without dimensionality reduction, Procrustes R$^2$ is computed on 15 dimensional PCA space.

\paragraph{Time-resolved trajectories.}
For trajectory metrics, each time step is treated as a separate observation. The tensor is reshaped to $(T \cdot S \cdot D, N)$, PCA is fitted on all valid (non-NaN) rows, and the result is reshaped back to $(SK, T, P)$ trajectories, where $P = 15$ PCA components. NaN-padded time steps remain NaN in the projected coordinates and are excluded from all trajectory metric computations. rRSA is computed on the full-dimensional space, other metrics on the PCA space.

\subsubsection{Static Decoding Manifold Metrics}
\label{app:dec_metrics}

Let $\hat{X}_\mathrm{1}, \hat{X}_\mathrm{2} \in \mathbb{R}^{SK \times N}$ be the time-averaged response matrices of two neural populations.

\textbf{k-NN Decoding Accuracy.}

$$
\mathrm{Acc}(\hat{X}) = \frac{1}{SK} \sum_{i=1}^{SK} \mathbf{1}\!\left[\hat{y}_i = y_i\right], 
$$
$$
\hat{y}_i = \underset{c}{\arg\max} \sum_{j \in NN_k(i)} \mathbf{1}[y_j = c],
$$

where $y_i$ is the stimulus label of point $i$ and $NN_k(i)$ is its leave-one-out $k$-nearest-neighbor set ($k{=}5$). Measures categorical separability.

\textbf{Representational Similarity Analysis (RSA).} Following \citet{kriegeskorteRepresentationalSimilarityAnalysis2008}, construct the representational dissimilarity matrix (RDM):

$$
\mathbf{D}^\mathrm{1}_{ij} = \bigl\|\bar{x}_i^\mathrm{1} - \bar{x}_j^\mathrm{1}\bigr\|_2, \qquad
\mathbf{D}^\mathrm{2}_{ij} = \bigl\|\bar{x}_i^\mathrm{2} - \bar{x}_j^\mathrm{2}\bigr\|_2,
$$

where $\bar{x}_i = \frac{1}{T}\sum_t X_{:,i,t}$ is the time-averaged population response to condition $i$. The RSA score is the Spearman rank correlation between the two upper-triangular distance vectors:

$$
\mathrm{RSA} = \rho_s\!\left(\mathrm{vec}(\mathbf{D}^\mathrm{1}),\; \mathrm{vec}(\mathbf{D}^\mathrm{2})\right) \in [-1, 1].
$$

RSA measures whether the stimulus-similarity geometry of the two populations match. RSA is sensitive only to ordinal structure, not absolute distances.

\textbf{Linear CKA (Centered Kernel Alignment).} Let $\mathbf{H} = \mathbf{I} - \frac{1}{m}\mathbf{1}\mathbf{1}^\top$ be the centering matrix ($m = S*K$):

$$
\mathrm{CKA}(\hat{X}_\mathrm{1},\hat{X}_\mathrm{2})
= \frac{\bigl\|\hat{X}_\mathrm{1}^\top \mathbf{H} \hat{X}_\mathrm{2}\bigr\|_F^2}
       {\bigl\|\hat{X}_\mathrm{1}^\top \mathbf{H} \hat{X}_\mathrm{1}\bigr\|_F
        \cdot \bigl\|\hat{X}_\mathrm{2}^\top \mathbf{H} \hat{X}_\mathrm{2}\bigr\|_F}.
$$

$\mathrm{CKA} \in [0,1]$; invariant to orthogonal transformations and isotropic scaling \citep{kornblithSimilarityNeuralNetwork2019, williamsGeneralizedShapeMetrics2022}.

\textbf{Procrustes $R^2$.} Let $\mathbf{Z}_\mathrm{1}, \mathbf{Z}_\mathrm{2} \in \mathbb{R}^{SK \times P}$ denote their top-$P = 15$ PCA projections (for subpopulations of $k < 3$ neurons, $P = k$; the smaller manifold is zero-padded to 15 dimensions before alignment). Now solve the orthogonal Procrustes problem over the $P$-dimensional PCA embeddings (after mean-centering and unit-norm scaling $\rightarrow \tilde{\mathbf{Z}}$):

$$
R^2_\mathrm{Proc} = 1 - d^*_\mathrm{Proc}, \qquad
d^*_\mathrm{Proc} = \min_{\mathbf{R} \in \mathcal{O}(P)} \bigl\|\tilde{\mathbf{Z}}_\mathrm{1} - \tilde{\mathbf{Z}}_\mathrm{2}\mathbf{R}\bigr\|_F^2.
$$

$R^2_\mathrm{Proc} \in [0,1]$; values near 1 indicate the two point clouds are nearly congruent after optimal rigid alignment \citep{williamsGeneralizedShapeMetrics2022}. Unlike CKA, Procrustes penalizes shape differences rather than alignment differences.

\subsubsection{Trajectory Metrics}

tRSA operates directly in native neural space without any dimensionality reduction. SPC, tProc and DSA each project their population's time-resolved responses independently onto $P=15$ PCA components (fitted per population via \texttt{compute\_decoding\_trajectories}). Let $\mathbf{z}_s^\mathrm{1}(t) \in \mathbb{R}^P$ and $\mathbf{z}_s^\mathrm{2}(t) \in \mathbb{R}^P$ denote the projected trajectory at time $t$ for stimulus condition $s$ (where applicable).

\textbf{Time-Resolved RSA (tRSA).} At each time bin $t$, compute instantaneous RDMs and their Spearman correlation:

$$
\mathbf{D}^\mathrm{1,2}_{ij}(t) = \bigl\|\mathbf{z}_i^\mathrm{1,2}(t) - \mathbf{z}_j^\mathrm{1,2}(t)\bigr\|_2, 
$$
$$
\mathrm{RSA}(t) = \rho_s\!\bigl(\mathrm{vec}(\mathbf{D}^\mathrm{1}(t)),\; \mathrm{vec}(\mathbf{D}^\mathrm{2}(t))\bigr).
$$

Summary scalar: $\overline{\mathrm{tRSA}} = \frac{1}{T}\sum_t \mathrm{RSA}(t)$.

\textbf{Speed Profile Correlation (SPC).}

$$
v_s^\mathrm{1,2}(t) = \bigl\|\mathbf{z}_s^\mathrm{1,2}(t{+}1) - \mathbf{z}_s^\mathrm{1,2}(t)\bigr\|_2,
$$
$$
\rho_s^\mathrm{speed} = \mathrm{Pearson}\!\left(v_s^\mathrm{1}(1{:}T{-}1),\; v_s^\mathrm{2}(1{:}T{-}1)\right).
$$

Summary scalar: $\overline{\mathrm{SPC}} = \frac{1}{SK}\sum_{s,k}\rho_{s,k}^\mathrm{speed}$. Captures whether the \emph{timing} of fast and slow phases of neural dynamics is preserved, even if absolute speeds differ.

\textbf{Trajectory Procrustes $R^2$ (tProc).} For each stimulus condition $s$, treat its $T$-step trajectory as a point cloud $\tilde{\mathbf{Z}}_s \in \mathbb{R}^{T \times P}$ (mean-centered and unit-norm-scaled) and apply orthogonal Procrustes alignment:

$$
R^2_{\mathrm{Proc},s}
= 1 - \min_{\mathbf{R} \in \mathcal{O}(P)} \bigl\|\tilde{\mathbf{Z}}_s^\mathrm{1} - \tilde{\mathbf{Z}}_s^\mathrm{2}\,\mathbf{R}\bigr\|_F^2.
$$

Summary scalar: $\overline{R^2_\mathrm{tProc}} = \frac{1}{SK}\sum_{s,k} R^2_{\mathrm{Proc},s,k}$. This is the direct temporal analogue of Proc.: it measures whether the trajectory \emph{shape} is preserved, not just the time-averaged endpoint.

\textbf{Dynamical Similarity Analysis (DSA).} Following \citet{ostrowGeometryComparingTemporal2023}, DSA fits a linear dynamical system to each population's PCA-reduced trajectory via Dynamic Mode Decomposition (DMD) and computes the Procrustes distance between the learned transition operators $\mathcal{A}_\mathrm{1}$ and $\mathcal{A}_\mathrm{2}$:

$$
\mathrm{DSA} = d_\mathrm{DSA}(\mathcal{A}_\mathrm{1},\, \mathcal{A}_\mathrm{2}),
$$

where $d_\mathrm{DSA}$ is the DSA distance. We report normalized (against the full population) z-scores of DSA values in comparison to 5 temporally shuffled baseline seeds. Unlike other trajectory metrics, which compare trajectories geometrically, DSA asks whether the two populations implement the \emph{same dynamical transformation} of stimuli over time (Appendix~\ref{app:dsa_method}).

\subsection{Encoding manifold construction}
\label{app:encoding_pipeline}

The encoding manifold is constructed following the three-stage pipeline of \citet{dyballaPopulationEncodingStimulus2024}: (1) nonnegative tensor factorization to embed neurons into a stimulus-response space \citep{williamsUnsupervisedDiscoveryDemixed2018}, (2) adaptive graph construction via IAN \citep{dyballaIANIteratedAdaptive2023}, and (3) diffusion maps \citep{coifmanDiffusionMaps2006} for dimensionality reduction.

\subsubsection{Preprocessing}
The raw $(N \times S \times K \times T)$ tensor is preprocessed as follows. First, responses are smoothed along the time axis with a 1D Gaussian kernel ($\sigma_{\mathrm{NTF}} = 3$ bins). Next, for datasets with two spatial frequencies (11 stimuli), per-neuron optimal spatial frequency is selected based on population-level response magnitude, reducing the stimulus dimension to $S = 6$. The $K = 8$ directional responses are concatenated along the time axis to form a single response vector per (neuron, stimulus) pair of length $K \times T$. Finally, each neuron's response is normalized by its relative firing rate (\texttt{relNorm} method: divide by each neuron's maximum response, then rescale by relative activation). The resulting 3-tensor has shape $(N \times S \times (K \cdot T))$.

\subsubsection{Nonnegative tensor factorization (NTF)}
The preprocessed tensor $\mathbf{T} \in \mathbb{R}^{N \times S \times (DT)}$ is decomposed into $R$ rank-1 nonnegative components \citep{williamsUnsupervisedDiscoveryDemixed2018}:
$$
\tilde{\mathbf{T}} = \sum_{r=1}^{R} \sigma^{\mathrm{NTF}}_r \, \mathbf{v}_r^{(1)} \circ \mathbf{v}_r^{(2)} \circ \mathbf{v}_r^{(3)},
$$
where $\mathbf{v}_r^{(1)}$ are neural factors, $\mathbf{v}_r^{(2)}$ stimulus factors, and $\mathbf{v}_r^{(3)}$ temporal response factors. Factor vectors are normalized to unit length with magnitudes absorbed by $\sigma^{\mathrm{NTF}}_r$. The decomposition is performed using the OPT method from Tensor Toolbox \citep{tensor_toolbox} with non-negativity constraints, run 50 times with different random initializations; the result with smallest reconstruction error is selected. The number of components $R$ (typically 8--17) is chosen per dataset based on the explained variance heuristic of \citet{dyballaPopulationEncodingStimulus2024}. Circular permutations of the directional axis are applied during decomposition to detect response patterns irrespective of preferred orientation.

\subsubsection{Neural encoding space}
Following the reformulation in \citet{dyballaPopulationEncodingStimulus2024}, the scaled neural factor matrix $\mathcal{N}_{\sigma_{\mathrm{NTF}}} = \mathcal{N} \mathbf{\Sigma_{\mathrm{NTF}}}$ (where $\mathcal{N} = \mathbf{X}^{(1)}$ and $\mathbf{\Sigma}_{\mathrm{NTF}} = \mathrm{diag}(\sigma^{\mathrm{NTF}}_1, \dots, \sigma^{\mathrm{NTF}}_R)$) places each neuron as a point in an $R$-dimensional stimulus-response space. Distances in this space reflect similarity in tuning and temporal dynamics. PCA is applied to $\mathcal{N}_{\sigma_{\mathrm{NTF}}}$ to reduce dimensionality; the number of retained components is chosen so that cumulative explained variance exceeds $80\%$.

\subsubsection{Outlier removal}
Before graph construction, $n_\text{far} = 2$ most-isolated and $n_\text{close} = 5$ most-similar neurons (by minimum pairwise Euclidean distance) are removed as outliers.

\subsubsection{Iterated Adaptive Neighborhoods (IAN)}
On the PCA-reduced neural encoding space, a weighted data graph is constructed using the IAN algorithm \citep{dyballaIANIteratedAdaptive2023}. IAN infers an adaptive local similarity kernel without requiring a fixed neighborhood size: it constructs the Gabriel graph, builds a multiscale Gaussian kernel, and iteratively prunes edges for consistency between discrete and continuous neighborhoods. Disconnected points identified by IAN are removed from subsequent analysis. The resulting sparse weighted adjacency matrix $\mathbf{K}$ encodes local similarities via locally tuned Gaussian kernels.

\subsubsection{Diffusion maps}
Diffusion maps \citep{coifmanDiffusionMaps2006} are applied to the IAN-weighted adjacency matrix $\mathbf{K}$. The matrix is normalized and symmetrized:
$$
\mathbf{d}_i = \sqrt{\textstyle\sum_j \mathbf{K}_{ij} + \epsilon}, \qquad
\mathbf{M}_s = \frac{\mathbf{K}}{\mathbf{d}\mathbf{d}^\top}.
$$
The spectral decomposition of $\mathbf{M}_s$ yields eigenvalues $\mu_0 \geq \mu_1 \geq \cdots$ and eigenvectors $\boldsymbol{\psi}_l$. We compute $L = 20$ eigenvectors with diffusion time $t = 1$ and $\alpha = 1$, producing diffusion coordinates $\Psi_t(i) = (\mu_l^t \boldsymbol{\psi}_l(i))_{l=0}^{L-1}$.

\subsubsection{MDS visualization}
For 3D visualization, metric MDS is applied to the pairwise squared Euclidean distances computed from the first diffusion coordinates: the Gram matrix $\mathbf{G} = -\frac{1}{2} \mathbf{H} \mathbf{D}^2 \mathbf{H}$ (where $\mathbf{H}$ is the centering matrix) is eigendecomposed and the leading 10 components are retained.

\subsection{Encoding Pipeline Robustness}
\label{app:enc_robustness}

To validate that encoding manifold topology is stable across hyperparameters, we rerun the full encoding pipeline on V1 and Retina under all combinations of CP rank $F \in \{10, 15, 20\}$ and neuron fraction $\in \{50\%, 75\%, 100\%\}$, yielding 18 conditions per dataset. The pipeline per condition is: tensor4d $\to$ CP (5 restarts, max 200 iterations) $\to$ IAN $\to$ diffusion maps (20 components, $\alpha=1$, $t=1$) $\to$ MDS (10 components) $\to$ HDBSCAN (min\_cluster\_size $= 25$, min\_samples $= 4$). Pairwise GW distances between all 18 embeddings (subsampled to 1200 points) are visualized as a distance heatmap and MDS projection (Figure~\ref{fig:enc_stability}).

\subsection{Population-size sweep (Section~\ref{sec:sweep})}
\label{app:sweep_method}

For each dataset, population fractions $f$ drawn from 14 preset values $\{0.01, 0.02, 0.05, 0.10, 0.15, 0.20, 0.30, 0.40, 0.50, 0.60, \\0.70, 0.80, 0.90, 1.00\}$, augmented by per-dataset absolute neuron counts $\{1, 2, 3, 5, 8, 13, 20\}$ (converted to fractions), are evaluated. At each fraction, $k = \lfloor f \cdot N \rfloor$ neurons are selected under six strategy types, each with high (top-$k$) and low (bottom-$k$) variants, yielding 11 distinct selection procedures:
\begin{enumerate}
    \item \textbf{Random}: uniform random without replacement (10 seeds; mean and standard deviation reported).
    \item \textbf{High/low curvature}: neurons ranked by RMS residual of a linear fit to the early Gaussian-smoothed response ($\sigma_{\mathrm{smooth}} = 2$ bins, first $T_\mathrm{early} = \min(40, T)$ time steps).
    \item \textbf{High/low stability}: neurons ranked by mean absolute temporal derivative over the last $T_\mathrm{early}$ time steps (lower = more stable).
    \item \textbf{High/low classifiability}: neurons ranked by variance of stimulus-averaged steady-state responses (last 10 time bins).
    \item \textbf{High/low OSI}: neurons ranked by orientation selectivity index; skipped for single-direction datasets (natural movie, $D=1$).
    \item \textbf{High/low PC contribution}: neurons ranked by the L2-norm of their loading on the first global principal component of the full tensor reshaped to $(S \cdot K \cdot T) \times N$.
\end{enumerate}

All eight metrics are computed for each subpopulation versus the full population.

\subsection{Encoding manifold region sampling}
\label{app:region_method}

For each dataset, two size-matched subpopulations are constructed by selecting the top-$k$ (high-PC condition) or bottom-$k$ (low-PC condition) neurons ranked by \textbf{PC contribution} --- the L2-norm of each neuron's loading on the first global principal component of the decoding tensor reshaped to $(S \cdot K \cdot T) \times N$ (strategy~6 in Appendix~\ref{app:sweep_method}). Three fractions are evaluated: $f \in \{0.01, 0.05, 0.10\}$, giving $k = \lfloor f N \rfloor$.

All eight decoding metrics are computed vs.\ the full population. GW similarity between the subpopulation encoding manifold (first 3 MDS dimensions, $\leq 1200$ points subsampled) and the full population is also computed.

The GW comparison panel (Figure~\ref{fig:regions5}J) pools GW values across all three fractions and four datasets (Retina, V1, VISp drifting gratings, VISp natural movie), yielding 12 matched pairs per condition (4 datasets $\times$ 3 fractions). Statistical significance is assessed with a Wilcoxon signed-rank test on matched pairs.

\subsection{Farthest Point Sampling (Section~\ref{sec:fps})}
\label{app:fps_method}

FPS constructs subpopulations with controlled topological coverage of the encoding manifold:

\begin{enumerate}
    \item \textbf{Seed selection.} The first seed is chosen uniformly at random from the encoding manifold embedding (first 3 MDS dimensions). Subsequent seeds are selected greedily by the farthest-point-sampling criterion: at each step, the neuron with the largest minimum Euclidean distance to all previously selected seeds is added. This produces $n$ seeds that maximally cover the manifold.
    \item \textbf{Nearest-neighbor expansion.} For each seed, its $m$ nearest neighbors in the full MDS embedding (all 10 dimensions) are added to the subpopulation, excluding neurons already selected. The resulting subpopulation contains at most $n \cdot (m + 1)$ neurons.
    \item \textbf{Conditions.} Two canonical conditions are compared:
    \begin{itemize}
        \item $(n{=}50, m{=}1)$: 50 widely spread seeds, each with one neighbor $\rightarrow$ 100 neurons spanning the manifold smoothly.
        \item $(n{=}10, m{=}9)$: 10 seeds, each with 9 neighbors $\rightarrow$ 100 neurons grouped into 10 tight clusters.
    \end{itemize}
    Both conditions produce subpopulations of $\approx 100$ neurons but with opposite encoding manifold topologies. Five random seeds are used for the initial seed selection: $\{42, 43, 44, 45, 46\}$. Mean and standard deviation across seeds are reported as central estimate and error bars in Figure~\ref{fig:fps}.
\end{enumerate}

All eight metrics are computed for each FPS subpopulation versus the full population.

\subsection{MNIST causal dissociation experiment (Section~\ref{sec:mnist})}
\label{app:mnist_method}

\subsubsection{Architecture}
We train a two-layer convolutional neural network on MNIST digit classification:
\begin{itemize}
    \item \texttt{Conv1}: $1 \to 16$ channels, $3 \times 3$ kernel $\to$ ReLU $\to$ MaxPool$(2)$ $\to$ $(B, 16, 13, 13)$
    \item \texttt{Conv2}: $16 \to 32$ channels, $3 \times 3$ kernel $\to$ ReLU $\to$ MaxPool$(2)$ $\to$ $(B, 32, 5, 5)$
    \item Flatten $\to$ $(B, 800)$ \quad (\textit{``neural population''})
    \item FC: $800 \to 10$ (logits)
\end{itemize}
Each of the $N = 800$ units in the flattened layer corresponds to a (channel, spatial position) pair in the 32 feature maps of \texttt{Conv2}. This layer is treated as the neural population for all encoding and decoding analyses.

\subsubsection{Training}
All models are trained with Adam (learning rate $10^{-3}$, batch size 256) for 5 epochs on the standard MNIST training set ($\sim 60\,000$ images). The standard MNIST test set ($10\,000$ images) is used for evaluation. We use 20 seeds for initialization to obtain statistically robust results.

\subsubsection{Baseline condition}
The baseline model is trained with cross-entropy loss only: $\mathcal{L} = \mathcal{L}_\text{CE}$.

\subsubsection{Clustering loss ($\mathcal{L}_\text{cluster}$)}
For each value of $\lambda \in \{0.001, 0.01, 0.1, 0.5, 1, 5, 10, 50\}$, the model is trained with:
$$
\mathcal{L} = \mathcal{L}_\text{CE} + \lambda \cdot \mathcal{L}_\text{cluster}.
$$

The clustering loss $\mathcal{L}_\text{cluster}$ is a supervised contrastive loss operating on the 800-dimensional activation profiles of neurons across the batch. For each neuron, its response vector across the $B$ samples in the current mini-batch serves as a high-dimensional feature. Neurons are labeled by their \emph{preferred stimulus} (the digit class eliciting the highest mean activation). The SupCon \citep{khoslaSupervisedContrastiveLearning2021} loss then pulls neurons with the same preferred stimulus together and pushes neurons with different preferences apart:
$$
\mathcal{L}_\text{cluster} = \frac{1}{|\mathcal{A}|} \sum_{i \in \mathcal{A}} \frac{1}{|P(i)|} \sum_{p \in P(i)} -\log \frac{\exp(\mathbf{z}_i \cdot \mathbf{z}_p / \tau)}{\sum_{j \neq i} \exp(\mathbf{z}_i \cdot \mathbf{z}_j / \tau)},
$$
where $\mathbf{z}_i = \mathbf{h}_i / \|\mathbf{h}_i\|$ is the $\ell_2$-normalized activation profile of neuron $i$ (a $B$-dimensional vector), $P(i) = \{j \neq i : \text{pref}(j) = \text{pref}(i)\}$ is the set of neurons sharing neuron $i$'s preferred digit, $\mathcal{A} = \{i : |P(i)| > 0\}$ is the set of anchor neurons with at least one positive, and $\tau = 0.07$ is the temperature.

This loss directly clusters the encoding manifold---neurons responding similarly to the same digit class are pulled toward shared prototypes---while the CE term ensures that task accuracy is maintained.

\subsubsection{Evaluation tensor}
For each trained model, a $(800 \times 10 \times 50 \times 1)$ response tensor is constructed: 50 randomly selected test images per digit class, with the 800-dimensional hidden activation recorded for each. The singleton fourth dimension ($T = 1$) is a placeholder for time, ensuring compatibility with the subpopulation analysis pipeline. Stimulus labels are $\mathbf{y} = (0, 0, \dots, 0, 1, 1, \dots, 9, 9)$ with 50 entries per class.

\subsubsection{Encoding manifold construction}
For each condition, the encoding manifold is built via the full CP~$\to$~IAN~$\to$~diffusion maps~$\to$~MDS pipeline (Appendix~\ref{app:encoding_pipeline}), with the following MNIST-specific settings:
\begin{itemize}
    \item CP rank $R = 10$, with $\texttt{NDIRS} = 1$ (no directional tuning). 5 random restarts of the CP optimizer.
    \item IAN is run on the $R$-dimensional neural encoding matrix with a small jitter ($\gamma = 10^{-4}$) added to break distance ties.
    \item Diffusion maps: $L = 20$ eigenvectors, $\alpha = 1$, $t = 1$.
    \item MDS: 5 components (first 2 used for 2D visualization).
\end{itemize}

\subsubsection{Encoding similarity (Gromov--Wasserstein)}
Encoding manifold similarity is quantified via the Gromov--Wasserstein (GW) distance via entropic regularization (Python Optimal Transport library, \texttt{ot.gromov.gromov\_wasserstein2}, \texttt{square\_loss}, max 10\,000 iterations, tolerance $10^{-4}$).

\subsubsection{Attribution analysis}
\label{app:mnist_attribution}

We use gradient$\times$activation attribution to quantify which neurons drive class-specific outputs.

\paragraph{FC layer (800 neurons).}
For each digit class $c$ and model seed, we pass $n = 200$ correctly classified test examples through the network, record the 800-dimensional hidden activation $\mathbf{h}$, and compute $\mathrm{attr}_{i,c} = |\partial \ell_c / \partial h_i| \cdot h_i$, where $\ell_c$ is the class-$c$ logit. Averaging over examples yields a $(10 \times 800)$ attribution matrix.

\paragraph{Conv2 layer (32 channels).}
The same procedure is applied to Conv2 pre-activations, spatially average-pooled over the $5 \times 5$ feature map, using $n = 300$ examples per class, yielding a $(10 \times 32)$ matrix.

\paragraph{Entropy.}
For class $c$, normalize absolute attributions: $\mathbf{p}_c = |\mathbf{a}_c| / \|\mathbf{a}_c\|_1$. Attribution entropy: $H_c = -\sum_i p_{c,i} \log p_{c,i}$; report $\bar{H} = \frac{1}{10} \sum_c H_c$. High entropy indicates distributed attribution across neurons; low entropy indicates concentration on a few.

\paragraph{Overlap.}
Mean cosine similarity between normalized attribution vectors across all 45 class pairs: $\bar{\mathrm{cos}} = \mathrm{mean}_{c \neq c'}(\hat{\mathbf{a}}_c \cdot \hat{\mathbf{a}}_{c'})$. High overlap indicates similar attribution patterns across classes (shared circuits); low overlap indicates class-specific circuits.

\paragraph{Assortativity.}
An undirected graph is built on neurons (FC: 800 nodes; Conv2: 32 channel nodes). Each node is labeled by its preferred digit: for FC neurons, the argmax of the mean tuning response; for Conv2 channels, the mode of neuron preferences within the channel. An edge $(u, v)$ exists iff the cosine similarity between L2-normalized weight vectors exceeds a threshold ($\theta_\text{FC} = 0.5$; $\theta_\text{Conv2} = 0.3$). Assortativity is computed via NetworkX. Positive values indicate that same-class neurons preferentially connect within class-specific pathways.

\subsubsection{Statistical significance}

We apply Wilcoxon rank-sum tests with Bonferroni-Holm correction to obtain statistically sound difference between the MNIST trained models.

\subsection{Dynamical Similarity Analysis (DSA)}
\label{app:dsa_method}

DSA \citep{ostrowGeometryComparingTemporal2023} is computed as follows. The response tensor is reduced to $P = \min(15, N)$ PCA components per population (randomized SVD, fitted independently for reference and subpopulation; NaN-padded frames replaced with zero). The DSA package fits a linear dynamical system via Dynamic Mode Decomposition (DMD) to each population's trajectories (shape $(SK, T, P)$) with $n_\text{delays} = \min(\lfloor T/2 \rfloor, 20)$ delay-embedding steps, adaptive rank $\mathrm{rank} = \min(P \cdot n_\text{delays},\, 2P)$, and Tikhonov regularization $\alpha_\text{DMD} = 10^{-2}$.

\textbf{Normalization.} A null distribution is obtained from $n_\text{shuf} = 10$ temporal permutations of the subpopulation trajectories, giving null mean $\bar{d}_\text{shuf}$ and std $\sigma_\text{shuf}$. The z-score $z = (\bar{d}_\text{shuf} - d) / \sigma_\text{shuf}$ is divided by the full-population self-z-score $z_\text{ref}$ (computed once per dataset) so that the full population scores exactly 1: $\mathrm{DSA} = \mathrm{clip}(z / z_\text{ref}, 0, 1)$. Implementation via the \texttt{dsa-analysis} package.

\subsection{Implementation details}
\label{app:implementation}

All experiments are implemented in Python using NumPy \citep{numpy}, SciPy \citep{SciPy}, scikit-learn \citep{scikit-learn}, and PyTorch \citep{pytorch}. The IAN algorithm uses the implementation of \citet{dyballaIANIteratedAdaptive2023} (v1.1.2). Diffusion maps use the \texttt{diffusionMapSparseK} function from the IAN package. Gromov--Wasserstein distances are computed with the Python Optimal Transport library \citep{flamaryPOTPythonOptimal2021}. DSA uses the \texttt{dsa-analysis} package. Visualizations use Matplotlib \citep{matplotlib}.

\subsection{Reproducibility}
\label{app:reproducibility}

All random seeds are fixed and reported: FPS sweep seeds = $\{42, 43, 44, 45, 46\}$, MNIST random initialization seeds = 0--19 for the population-size sweep. Code is available at \url{https://github.com/JohannesBertram/FNN_Manifolds}, and we provide an interactive \href{https://johannesbertram.github.io/FNN_Manifolds/index.html}{Neural Manifold Explorer} tool.

\end{document}